\documentclass{l4dc2027}

\title[Behavioral Denoising]{Behavioral Denoising: A New Perspective on Nonlinear Behavioral Control Inspired by Closed-Form Diffusion Policies}
\usepackage{times}
\usepackage{mathtools}
\usepackage{bm}
\usepackage{booktabs}
\usepackage{array}
\usepackage{xcolor}
\usepackage{wrapfig}
\usepackage{multirow}
\usepackage{capt-of} 

\SetAlCapHSkip{0pt}

\newcommand{\R}{\mathbb{R}}

\newcommand{\I}{\mathbb{I}}
\newcommand{\E}{\mathbb{E}}

\newcommand{\N}{\mathcal{N}}
\newcommand{\KL}{\mathrm{D}_{\mathrm{KL}}}
\newcommand{\Cset}{\mathcal{C}}
\newcommand{\Proj}{\mathcal{P}}
\newcommand{\mthd}[1]{\texttt{#1}}
\newcommand{\hist}{\mathrm{hist}}

\newcommand{\eff}{\mathrm{eff}}
\newcommand{\smp}{\mathrm{s}}
\newcommand{\one}{\mathbf{1}}
\newcommand{\zb}{\bar z}

\newcommand{\methodeq}[2]{\makebox[0pt][r]{\mthd{#1}\quad}\boxed{#2}}

\DeclareMathOperator*{\argmin}{argmin}

\DeclareMathOperator{\conv}{conv}
\DeclareMathOperator{\ri}{ri}
\DeclareMathOperator{\Cov}{Cov}

\author{%
 \Name{Lucas Y. Jia} \Email{yixuany@mit.edu}
 \AND
 \Name{John Z. Zhang} \Email{jzhang3@mit.edu}   
 \AND
 \Name{Zachary Manchester} \Email{zacm@mit.edu} \\
 \addr Massachusetts Institute of Technology
}

\begin{document}

\maketitle
\vspace{-1.5cm}
\begin{abstract}%
Diffusion policies have become a dominant approach to imitation learning in robotics. Recent work suggests that, in the low-data regime, they largely retrieve or combine the demonstrated actions, which the closed-form diffusion policy does explicitly. This closely resembles behavioral control, where future inputs are formed as linear combinations of recorded trajectories. 
In this work, we make this connection precise: we show that the weights with which the closed-form diffusion policy combines the demonstrated actions are the unique solution of a regularized behavioral control problem.
Building on this connection, we propose an algorithm, termed {\it $\chi^2$-projected behavioral denoising}, which is a training-free method that combines nearby trajectory data with a limit on extrapolation when the data offer little support.
It shows encouraging results on four nonlinear control tasks compared to recent nonlinear behavioral control methods. We hope that this connection offers a useful perspective for understanding both data-driven control and diffusion policies.
Code and more visualizations can be found at \url{https://github.com/RoboticExplorationLab/behavioral-denoising}.
\end{abstract}




\section{Introduction}\label{sec:intro}

\begin{wrapfigure}{r}{0.45\textwidth}   
  \centering
  \vspace{-12pt}                         
  \includegraphics[width=0.43\textwidth]{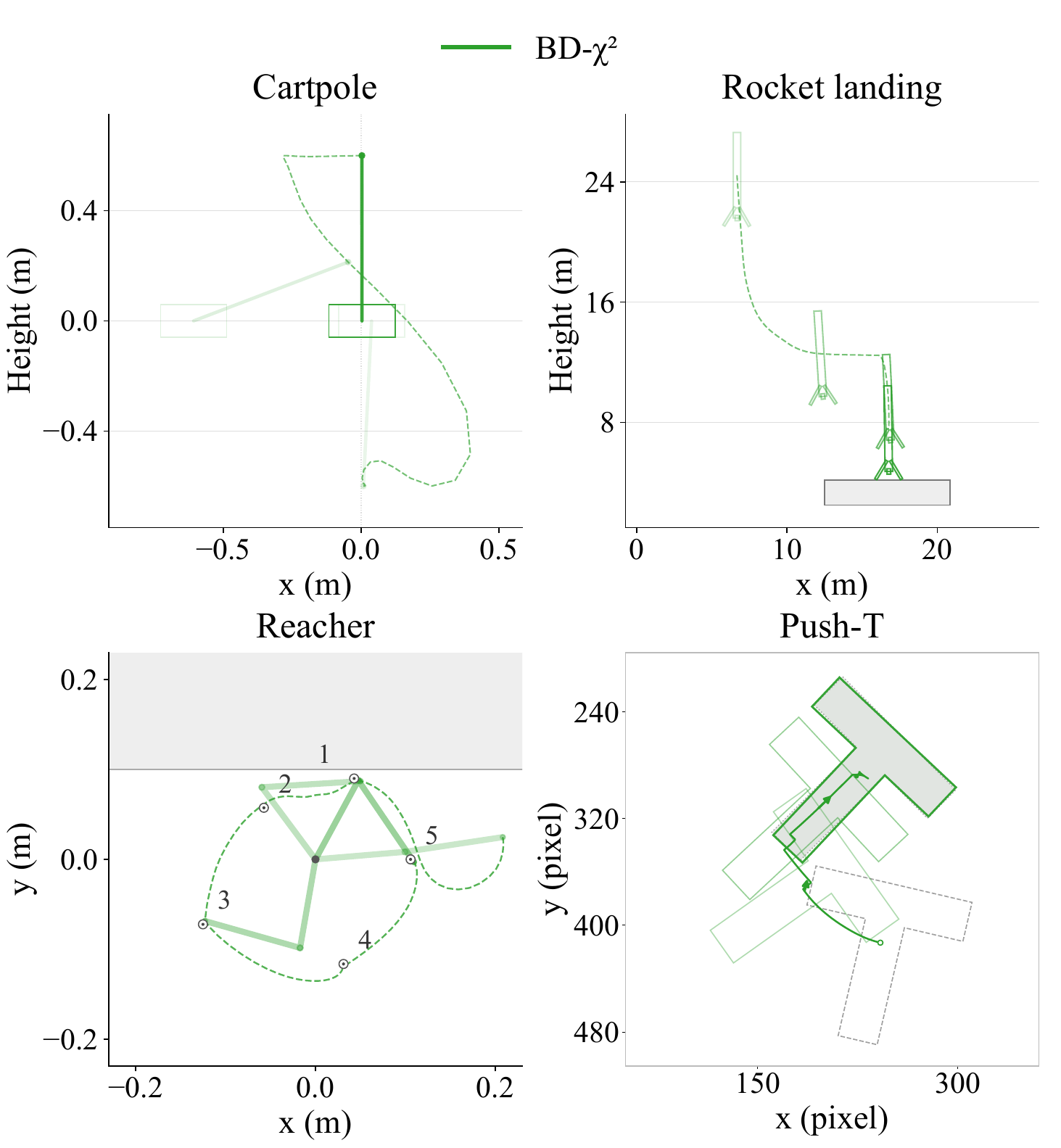}
  \caption{Examples of closed-loop trajectories of \mthd{BD-$\chi^2$}.}
  \label{fig:bd_chi2_demo}
  \vspace{45pt}                         
\end{wrapfigure}

Diffusion policies \citep{janner2022planning,chi2023diffusion}, which parametrize a control policy by a diffusion model \citep{sohl2015deep,ho2020denoising,song2021score}, have become a dominant approach to imitation learning in robotics \citep{chi2023diffusion,ze2024dp3,octo2024}. 
Given an observation history, such as the most recent camera images, a diffusion policy draws a noise sample that represents a noisy sequence of actions and denoises it step by step into a clean sequence, conditioning each step on the observation history. The first action, or the first few, is then applied, a new observation is made, and the process repeats.

\begin{figure}[!t]
\centering
\includegraphics[width=\textwidth]{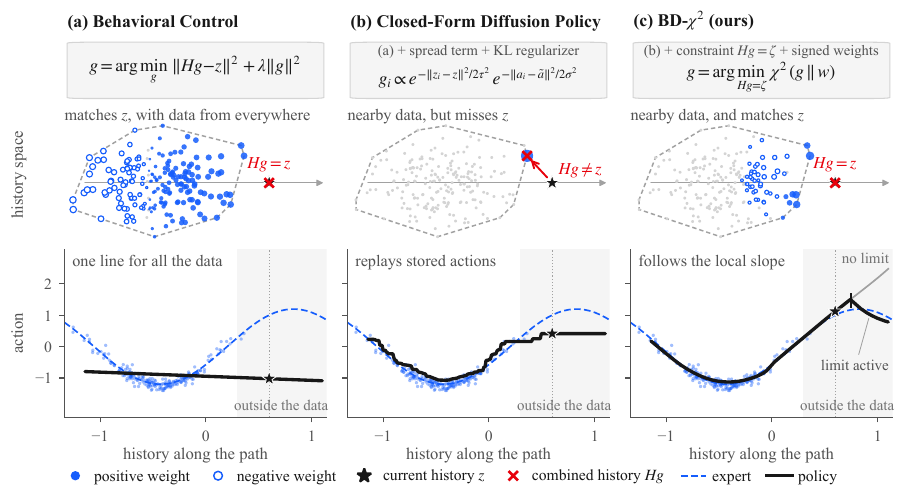}
\caption{Behavioral control, the closed-form diffusion policy, and \mthd{BD-$\chi^2$} on toy data, at a history outside the data (top) and along a path leaving the data (bottom). (a) Behavioral control reproduces $z$ but combines data from everywhere. (b) The closed-form diffusion policy (Proposition~\ref{prop:dbc}) combines only nearby data but does not reproduce $z$. (c) \mthd{BD-$\chi^2$} does both and follows the expert's local slope, up to a limit far from the data.}
\label{fig:fig1}
\vspace{-1cm}
\end{figure}

Why diffusion models work so well is not yet fully understood, but recent studies offer a partial explanation: With the exact score function of a finite training set, which is what perfect training would learn, a diffusion model only reproduces its training samples \citep{scarvelis2023closed}. New samples arise from the approximation error of the score, and replacing this error by an explicit smoothing of the closed-form score yields a model that provably outputs weighted combinations of the training samples \citep{scarvelis2023closed}. Studies of diffusion policies point in a similar direction: In the low-data regime, they appear to largely retrieve or combine the demonstrated actions \citep{he2026demystifying}, and a closed-form diffusion policy, which requires no training, obtains competitive results \citep{mishra2026training}. These findings suggest that, at least in some regimes, the action of a diffusion policy can be viewed as a weighted combination of the demonstrated actions.

Interestingly, forming new inputs as combinations of recorded trajectories is also the central idea of behavioral control, a data-driven control paradigm that designs controllers directly from measured input-observation data without explicitly identifying a model~\citep{hou2013model,recht2019tour}.
For a linear time-invariant system, the fundamental lemma of \citet{willems2005note} states that one sufficiently exciting trajectory spans all trajectories of the system, so a matrix of recorded data can serve as a nonparametric model for control \citep{markovsky2021behavioral}. Behavioral control uses this representation to predict and choose future inputs as combinations of recorded trajectories \citep{coulson2019data,de2020formulas,berberich2021data,markovsky2023data}.

The theory of behavioral control is exact only for linear systems, and much effort has gone into extending it to nonlinear systems. Data-enabled predictive control (DeePC) \citep{coulson2019data} showed that suitable regularization gives good tracking on a simulated nonlinear quadrotor, and was later demonstrated on quadcopter hardware \citep{elokda2021data}. Its regularizers were subsequently explained as robustness to uncertainty in the data \citep{coulson2019regularized,coulson2022distributionally,huang2023robust} and as a relaxation between direct and indirect data-driven control \citep{dorfler2023bridging}. Other extensions either lift the data into a higher-dimensional space in which the behavior is approximately linear \citep{lian2021nonlinear,huang2024robust}, or select the part of the dataset near the current operating point, which implicitly linearizes the behavior locally \citep{berberich2022linear,naf2025choose}. In all of them, the inputs are still formed as combinations of recorded trajectories.

Diffusion policies and behavioral control thus share the idea of producing new inputs as combinations of recorded ones, but they come from different communities and are studied with different tools. In this work, we study this resemblance, with the aim of providing a different perspective on designing behavioral control algorithms for nonlinear systems.
Our contributions are as follows:
\begin{itemize}
\setlength{\itemsep}{3pt}
\setlength{\parskip}{3pt}
\item We show that a regularized behavioral control problem, which we call {\it behavioral denoising}, has the weights of the closed-form diffusion policy of \citet{mishra2026training} as its unique solution (Section~\ref{sec:dbc}). This connection places the two at opposite ends of a trade-off between \emph{consistency}, i.e., reproducing the observed history, and \emph{locality}, i.e., combining only nearby data (Section~\ref{sec:limits}).
\item Building on this connection, we derive a $\chi^2$ projection of the behavioral denoiser that, like the closed-form diffusion policy, combines only nearby data and, like behavioral control, reproduces the observed history, in closed form (Section~\ref{sec:consistency}).
\item With a few practical modifications, we turn this projection into a training-free algorithm, $\chi^2$-projected behavioral denoising (\mthd{BD-$\chi^2$}, Section~\ref{sec:method}). It shows encouraging results on four nonlinear control tasks (Section~\ref{sec:experiment}).
\end{itemize}

\vspace{-0.6cm}
\section{Notation and Background}\label{sec:setup}

Throughout, $\one$ denotes the vector of all ones and $\I$ the identity matrix, with sizes clear from the context. For a vector $v$, $\mathrm{diag}(v)$ is the diagonal matrix with the entries of $v$ on its diagonal. $\conv\{x_1,\dots,x_N\}$ is the convex hull of the points $x_i$, i.e., the set of all $\sum_ig_ix_i$ with $g\ge0$ and $\one^\top g=1$, and $\ri\conv\{x_1,\dots,x_N\}$ is its relative interior, which consists of the combinations with $g>0$. $\N(x;m,\Sigma)$ is the density at $x$ of the Gaussian distribution with mean $m$ and covariance $\Sigma$. Sums without limits run over all indices, and $g_i\propto c_i$ means $g_i=c_i/\sum_jc_j$.

\vspace{-0.3cm}
\subsection{Notation for Demonstrations}\label{sec:notation}
Let $u^{(t)}\in\R^{n_u}$ and $y^{(t)}\in\R^{n_y}$ denote the input and the observation at time $t$. From the demonstrated trajectories (e.g. collected from an MPC controller), we extract $N$ windows. A window $i$, taken at time $t$ of a demonstration, stores a \emph{history} $z_i\in\R^{n_z}$, which stacks the inputs $u^{(t-T_\hist)}$ to $u^{(t-1)}$ and the observations $y^{(t-T_\hist+1)}$ to $y^{(t)}$, and an \emph{action}\footnote{Note that an action at time $t$ can be the control input $u^{(t)}$, or a chunk of the next $T_c$ inputs.} $a_i\in\R^{n_a}$ that the expert (e.g. the MPC controller) computed at time $t$. At run time, the controller observes the \emph{current history} $z$, a vector in $\R^{n_z}$ built in the same way, and returns an action. The windows are collected column-wise in:
\begin{equation}
H\coloneqq\begin{bmatrix}z_1&\cdots&z_N\end{bmatrix},\qquad
A\coloneqq\begin{bmatrix}a_1&\cdots&a_N\end{bmatrix}.
\label{eq:matrices}
\end{equation}

Every policy in this paper returns a combination $Ag$ of the stored actions, with weights $g\in\R^N$ that sum to one. Note that the weights also combine the stored histories into $Hg$. Let $\Delta_N\coloneqq\{g\in\R^N:g\ge0,\ \one^\top g=1\}$ and $\Delta_N^\circ\coloneqq\{g\in\Delta_N:g_i>0\ \forall i\}$. We call
$\Cset(\zeta)\coloneqq\{g\in\R^N:\ \one^\top g=1,\ Hg=\zeta\}$
the \emph{consistency set} at $\zeta$, and $g\in\Cset(z)$ \emph{consistent}: its combined history reproduces the current history $z$ exactly. For $w\in\Delta_N^\circ$ we define:
\begin{equation}
W_w\coloneqq\mathrm{diag}(w)-ww^\top,\qquad
S_w\coloneqq HW_wH^\top,\qquad
K_w\coloneqq AW_wH^\top S_w^{-1},
\label{eq:moments}
\end{equation}
where we assume $S_w\succ0$. $S_w$ is the $w$-weighted covariance of the histories around their mean $Hw$, and $K_w$ is the slope of the $w$-weighted least-squares fit of the actions on the histories, which can be interpreted as an estimate of the feedback gain of the demonstrator (Appendix~\ref{app:Kw}). The subscript $w$ is dropped when clear.

\vspace{-0.2cm}
\subsection{Closed-Form Diffusion Policy}\label{sec:cfd}\label{sec:cfdp}
For a prior weight $\pi\in\Delta_N^\circ$ and bandwidths $\tau,\sigma>0$, we define the weights:
\begin{equation}
g^\sigma_i(z,\tilde a)\coloneqq\frac{\pi_i\,e^{-\phi_i(z,\tilde a)}}{\sum_j\pi_j\,e^{-\phi_j(z,\tilde a)}},\qquad
\phi_i(z,\tilde a)\coloneqq\frac{\|z_i-z\|^2}{2\tau^2}+\frac{\|a_i-\tilde a\|^2}{2\sigma^2},
\label{eq:cfdp_weights}
\end{equation}
and the limit of $g^\sigma_i(z,\tilde a)$ as $\sigma\to\infty$, $q^\tau_i(z)\propto\pi_i\,e^{-\|z_i-z\|^2/(2\tau^2)}$, which depends on the history alone. 
Note that $g^\sigma(z,\tilde a)$ is the posterior over the windows given the noisy pair $(z,\tilde a)$, when the stored histories and actions are noised with Gaussian noise of standard deviations $\tau$ and $\sigma$ (in which case the noisy pair has density $\tilde p(z,\tilde a;\tau,\sigma)\coloneqq\sum_j\pi_j\,\N(z;z_j,\tau^2\I)\,\N(\tilde a;a_j,\sigma^2\I)$). The posterior mean $(Hg^\sigma,Ag^\sigma)$ is the denoiser that minimizes the denoising score-matching loss \citep{karras2022elucidating}, and by Tweedie's formula \citep{efron2011tweedie} it moves the noisy pair along the score of $\tilde p$:
\begin{equation}
Hg^\sigma(z,\tilde a)-z=\tau^2\,\nabla_z\log\tilde p(z,\tilde a;\tau,\sigma),\qquad
Ag^\sigma(z,\tilde a)-\tilde a=\sigma^2\,\nabla_{\tilde a}\log\tilde p(z,\tilde a;\tau,\sigma).
\label{eq:histresidual}
\end{equation}
The closed-form diffusion policy of \citet{mishra2026training} denoises a noisy action $\tilde a$ with the stored actions $a_i$ as data points, and additionally weights each window by how close its history is to the current history $z$. 
In our notation, its denoiser at noise level $\sigma>0$ is $D(z,\tilde a;\sigma)\coloneqq Ag^\sigma(z,\tilde a)$.


\subsection{Behavioral Control}\label{sec:bc}
Behavioral control \citep{willems2005note,coulson2019data} represents the future inputs at the current observed history by a combination of the stored ones that approximately reproduces the history. One way to formulate the problem, following \citet{adang2026training}, is:
\begin{equation}
g^{\mathrm{bc}}\in\argmin_{\one^\top g=1}\ \frac12\|Hg-z\|^2+\lambda\,\Omega(g),\quad u^{\mathrm{bc}}\coloneqq A\,g^{\mathrm{bc}},
\label{eq:bc}
\end{equation}
with a regularizer $\Omega$ on the weights $g$, which may be negative.
We use this formulation as a starting point for the motivation and development of the proposed algorithm. In practice, one would add extra cost terms (e.g. tracking desired observations) as done in DeePC \citep{coulson2019data}.


\section{Connecting Behavioral Control to Closed-form Diffusion Policies}\label{sec:dbc}
 We first give an overview of a unifying formulation that captures all the methods discussed in this paper (Section~\ref{sec:template}).
In this section, we show that one can regularize \eqref{eq:bc} such that its unique closed-form solution matches the expression of the closed-form diffusion policy \eqref{eq:cfdp_weights} (Section~\ref{sec:objective}). We then show that naively applying the resulting policy has two important limitations (Section~\ref{sec:limits}).

\subsection{A Unifying Formulation}\label{sec:template}
Both the closed-form diffusion policy and policies constructed using behavioral control return controls in the form of $Ag$, and differ mainly in how they compute the weight $g$. The weights of every policy in this paper have the form:
\begin{equation}
g=\argmin_{\one^\top g=1}\ D_f(g\,\|\,w)+\frac{1}{2\lambda}\|Hg-\zeta\|^2,
\label{eq:template}
\end{equation}
with \emph{reference weights} $w\in\Delta_N^\circ$, a divergence $D_f$, a target history $\zeta$, and $\lambda\in[0,\infty]$: for $\lambda=\infty$ the history term is dropped and $g=w$, and for $\lambda=0$ it is the constraint $g\in\Cset(\zeta)$. We use two divergences: the KL divergence $\KL(g\,\|\,w)$, defined for $g\in\Delta_N$, which keeps the action in the convex hull of the stored actions, and the $\chi^2$ divergence $\chi^2(g\,\|\,w)\coloneqq\sum_i(g_i-w_i)^2/w_i$, which allows negative weights and yields a closed-form minimizer (Section~\ref{sec:signed}). Behavioral control \eqref{eq:bc} with the quadratic regularizer $\Omega(g)=g^\top\mathrm{diag}(w)^{-1}g=\chi^2(g\,\|\,w)+1$ yields \eqref{eq:template} with $\zeta=z$. Table~\ref{tab:template} in the appendix lists the policies of this paper as choices of $(w,D_f,\lambda,\zeta)$. 


\subsection{Behavioral Denoising}\label{sec:objective}



For $g\in\Delta_N$ and a matrix $M\in\{H,A\}$ with columns $m_i$, let $V^M(g)\coloneqq\sum_ig_i\|m_i-Mg\|^2$ be the \emph{spread} of the columns around their combination. Expanding the square around $Mg$ gives, for any $x$ of appropriate dimension (Appendix~\ref{app:split}),
\begin{equation}
\|Mg-x\|^2+V^M(g)=\sum_ig_i\|m_i-x\|^2.
\label{eq:split}
\end{equation}
We consider the following problem, which we refer to as the \emph{behavioral denoising problem}:
\begin{equation}
    \min_{g\in\Delta_N}\ \frac{\|Hg-z\|^2+V^H(g)}{2\tau^2}+\frac{\|Ag-\tilde a\|^2+V^A(g)}{2\sigma^2}+\KL(g\,\|\,\pi).
\label{eq:dbc}
\end{equation}
Compared with \eqref{eq:bc}, the weights are now restricted to the simplex, the reconstruction term $\|Hg-z\|^2$ is accompanied by the spread $V^H(g)$ \eqref{eq:split}, and analogous terms $\|Ag-\tilde a\|^2$ and $V^A(g)$ are added for a noisy action $\tilde a$. The regularizer here is the KL divergence to a prior weight $\pi$. The bandwidths $\tau,\sigma>0$ set the scales on which histories and actions are compared.


\begin{proposition}\label{prop:dbc}
Let $\tau,\sigma>0$ and $\pi\in\Delta_N^\circ$. For any history $z$ and noisy action $\tilde a$, \eqref{eq:dbc} has a unique minimizer, and it is the weight vector $g^\sigma(z,\tilde a)$ of the closed-form diffusion policy \eqref{eq:cfdp_weights}.
\end{proposition}

\begin{proof}
Write $\phi_i$ for $\phi_i(z,\tilde a)$ and $g^\sigma$ for $g^\sigma(z,\tilde a)$, and let $Z\coloneqq\sum_j\pi_je^{-\phi_j}$. By \eqref{eq:cfdp_weights}, $g^\sigma_i=\pi_ie^{-\phi_i}/Z$, i.e., $\phi_i=\log(\pi_i/g^\sigma_i)-\log Z$. By \eqref{eq:split}, applied with $(M,x)=(H,z)$ and with $(M,x)=(A,\tilde a)$, the first two terms of \eqref{eq:dbc} equal $\sum_ig_i\phi_i$. Substituting $\phi_i$ and using $\sum_ig_i=1$, the objective of \eqref{eq:dbc} becomes, for every $g\in\Delta_N$,
\begin{equation}
\sum_ig_i\phi_i+\KL(g\,\|\,\pi)
=\sum_ig_i\log\frac{\pi_i}{g^\sigma_i}+\sum_ig_i\log\frac{g_i}{\pi_i}-\log Z
=\KL(g\,\|\,g^\sigma)-\log Z.
\label{eq:dbc_kl}
\end{equation}
$Z$ does not depend on $g$, and $\KL(g\,\|\,g^\sigma)\ge0$ with equality if and only if $g=g^\sigma$ (Gibbs' inequality). Hence $g^\sigma$ is the unique minimizer.
\end{proof}


\noindent
From here on, we call $g^\sigma(z,\tilde a)$ the \emph{behavioral denoiser}.



In terms of \eqref{eq:template}, the closed-form diffusion policy uses the reference $w=g^\sigma(z,\tilde a)$ with $\lambda=\infty$, and by Proposition~\ref{prop:dbc} this reference is itself the solution of a behavioral control problem with a spread term.

\vspace{-0.2cm}
\subsection{Behavioral Denoising as a Policy and Its Limitations}\label{sec:limits}\label{sec:policy}\label{sec:spread}



The denoiser $D(z,\tilde a;\sigma)=Ag^\sigma(z,\tilde a)$ of \eqref{eq:cfdp_weights} gives two policies, which we call behavioral denoising (BD) policies: one averages the stored actions (\mthd{BD-mean}), and the other samples one of them (\mthd{BD-sample}).
\mthd{BD-mean} lets $\sigma\to\infty$, where $g^\sigma\to q^\tau(z)$:
\begin{equation}
\methodeq{BD-mean}{u_{\mathrm{mean}}(z)\coloneqq Aq^\tau(z)=\sum_iq^\tau_i(z)\,a_i}
\label{eq:mean}
\end{equation}
It returns a weighted average of the stored actions, in which each window is weighted by how close its history is to the current history $z$.
\mthd{BD-sample}, the closed-form diffusion policy of \citet{mishra2026training}, holds $z$ fixed and integrates the probability-flow ODE \citep{karras2022elucidating}
\begin{equation}
\methodeq{BD-sample}{\frac{d\tilde a}{d\sigma}=\frac{\tilde a-D(z,\tilde a;\sigma)}{\sigma}}
\label{eq:ode}
\end{equation}
from $\tilde a=\sigma_{\max}\xi$, $\xi\sim\N(0,\I)$, to $\sigma=0$, and returns the endpoint $a^\smp(z,\xi)$. It returns approximately one stored action, chosen roughly with probability $q^\tau_i(z)$ (Appendix~\ref{app:flow_endpoint}).

To assess different policies, let $\kappa$ be a smooth feedback policy that describes the expert, i.e., the action the expert would apply at each history.\footnote{If the expert responds to the same history in several distinct ways (e.g., passing an obstacle on the left or on the right), $\kappa$ describes one of them, and the windows of the others count as deviations $\eta_i$ in Proposition~\ref{prop:approx}.}

\begin{proposition}\label{prop:approx}
Let $\kappa:\mathcal Z\to\R^{n_a}$ be differentiable on an open convex set $\mathcal Z$ containing $z$ and the $z_i$, with $\|\nabla\kappa\|\le L$ and $\nabla\kappa$ $B$-Lipschitz on $\mathcal Z$, and let $\eta_i\coloneqq a_i-\kappa(z_i)$ be the deviation of window $i$'s action from $\kappa$. Then, for every $\zeta\in\mathcal Z$ and $g\in\Cset(\zeta)$,
\[
\Big\|Ag-\kappa(z)\Big\|\le L\,\|z-\zeta\|+\frac B2\sum_i|g_i|\,\|z_i-\zeta\|^2+\Big\|\sum_ig_i\eta_i\Big\|.
\]
\end{proposition}

The first term is a \emph{first-order error} from reproducing $\zeta$ instead of $z$, which vanishes for consistent weights. The second is a \emph{curvature error}, which grows with the spread of the combined histories around $\zeta$. The third is a \emph{noise error}, the weighted average of the deviations $\eta_i$ (Appendix~\ref{app:prop_approx}).

The spread term determines where behavioral denoising sits in this bound. By \eqref{eq:split}, the first two terms of \eqref{eq:dbc} equal $\sum_ig_i\big(\|z_i-z\|^2/(2\tau^2)+\|a_i-\tilde a\|^2/(2\sigma^2)\big)$. They ask every window that is used to have its history close to $z$ and its action close to $\tilde a$ (\emph{locality}), which keeps the curvature error small. Nothing, however, asks the combination $(Hg,Ag)$ to reproduce $(z,\tilde a)$ (\emph{reconstruction}). By \eqref{eq:histresidual}, the combined history is $Hg^\sigma=z+\tau^2\nabla_z\log\tilde p(z,\tilde a;\tau,\sigma)$, a denoised version of $z$ pulled toward where nearby windows are dense, so a first-order error remains. Dropping the spread term reverses the trade-off: the combination reproduces $(z,\tilde a)$, but windows from anywhere in the data set are combined and the curvature error grows (Appendix~\ref{app:spread}).

Moreover, since the weights lie in $\Delta_N$, the combined history lies in $\conv\{z_1,\dots,z_N\}$ and the action in $\conv\{a_1,\dots,a_N\}$. When $z$ lies outside the first hull, no such weights can remove the first-order error, and the action cannot follow the expert beyond the demonstrated actions. Behavioral denoising therefore has two immediate limitations as a policy: it does not reproduce the current history, and it cannot extrapolate (see Appendix~\ref{app:limits} for more discussion).

\section{The $\chi^2$-Projected Behavioral Denoising Algorithm}\label{sec:consistency}

In this section, we propose both theoretical and practical fixes to the behavioral denoiser, which leads to the algorithm that we call the {\it $\chi^2$-projected behavioral denoising} (\mthd{BD-$\chi^2$}).

\subsection{Consistency as a Constraint}\label{sec:constraint}

A direct way to ask for consistency without giving up locality is to keep behavioral denoising and add the constraint $Hg=z$. By \eqref{eq:dbc_kl}, on $\Delta_N$ the objective of \eqref{eq:dbc} equals $\KL\big(g\,\|\,g^\sigma(z,\tilde a)\big)$ up to a constant, so the constrained problem projects the behavioral denoiser onto the consistency set in KL divergence:
\begin{equation}
\methodeq{BD-KL}{g^{\mathrm c}(z,\tilde a;\sigma)=\argmin_{g\in\Delta_N\cap\Cset(z)}\KL\big(g\,\|\,g^\sigma(z,\tilde a)\big)}
\label{eq:klproj}
\end{equation}
The minimizer exists and is unique whenever $z\in\conv\{z_1,\dots,z_N\}$ (Proposition~\ref{prop:klproj} in Appendix~\ref{app:sec3}). Using $g^{\mathrm c}$ in place of $g^\sigma$ in the sampling ODE \eqref{eq:ode} gives \mthd{BD-KL}, whose weights are consistent at every noise level, so the first-order error of Proposition~\ref{prop:approx} vanishes. However, its weights are nonnegative, so its action stays in $\conv\{a_1,\dots,a_N\}$. When $z$ lies outside $\conv\{z_1,\dots,z_N\}$, no such weights reproduce it, and \eqref{eq:klproj} has no solution.

\vspace{-0.3cm}
\subsection{Extrapolating with Signed Weights}\label{sec:signed}



To extrapolate, we must allow some weights to be negative, which the KL divergence forbids. However, its second-order expansion, the $\chi^2$ divergence of Section~\ref{sec:template}, is defined for signed weights: for $g$ near $w\in\Delta_N^\circ$ with $\one^\top g=1$,
\begin{equation}
\KL(g\,\|\,w)=\tfrac12\chi^2(g\,\|\,w)+O\big(\|g-w\|^3\big)
\label{eq:chi2}
\end{equation}%
(Appendix~\ref{app:chi2}; see also \citet{polyanskiy2025information}).

Replacing the divergence in \eqref{eq:klproj} by $\chi^2$ and dropping $g\ge0$ gives \eqref{eq:template} with $\lambda=0$ and turns it into a weighted minimum-norm problem, since $\chi^2(g\,\|\,w)=\sum_i(g_i^2/w_i-2g_i+w_i)=\sum_ig_i^2/w_i-1$ whenever $\one^\top g=1$.
Then, for a general weight vector $w\in\Delta_N^\circ$ and target history $\zeta$, we have the following problem:
\begin{equation}
\Proj_w(\zeta)\coloneqq\argmin_{g\in\Cset(\zeta)}\ \sum_{i=1}^N\frac{g_i^2}{w_i}.
\label{eq:proj}
\end{equation}
\noindent
Under a mild condition, this problem has a unique solution.
\begin{proposition}\label{prop:proj}
Let $w\in\Delta_N^\circ$ with $S_w\succ0$. Then $\Proj_w(\zeta)$ is unique,
\[
\Proj_w(\zeta)=w+W_wH^\top S_w^{-1}(\zeta-Hw),
\]%
and
\begin{equation}
A\,\Proj_w(\zeta)=Aw+K_w(\zeta-Hw).
\label{eq:projaction}
\end{equation}%
\end{proposition}

\noindent
The proof is given in Appendix~\ref{app:proj}.

The resulting consistent action $A\,\Proj_w(\zeta)=Aw+K_w(\zeta-Hw)$ \eqref{eq:projaction} starts from the weighted mean action $Aw$ and moves along the fitted slope $K_w$ toward $\zeta$. It is the $w$-weighted least-squares line through the windows, evaluated at $\zeta$.
With $\zeta=z$ it has no first-order error, and its curvature error is small when $w$ puts its weight on windows whose histories are close to $z$. With $w=q^\tau(z)$, the weights of \mthd{BD-mean}, we call it \mthd{BD-mean-$\chi^2$}:
\begin{equation}
\methodeq{BD-mean-$\chi^2$}{u^{\chi^2}_{\mathrm{mean}}(z)\coloneqq A\,\Proj_{q^\tau(z)}(z)=Aq^\tau+K_{q^\tau}(z-Hq^\tau)}
\label{eq:meanchi2}
\end{equation}%

\subsection{Practical Fixes}\label{sec:fixes}\label{sec:method}

In practice, the consistent action $A\Proj_w(\zeta)$ of Section~\ref{sec:signed} is only as good as the slope $K_w$, which is fitted to noisy data.
Windows whose actions do not follow the local trend of the others can pull the slope away from the expert's, and far from the data the noise in the actions is amplified in the extrapolated action.
We address these with several practical fixes, which lead to our proposed algorithm: $\chi^2$-projected behavioral denoising (\mthd{BD-$\chi^2$}).

\noindent
{\bf Fitting the line robustly.}\label{sec:select} We weight each window by its residual from a line $\zeta\mapsto c+K(\zeta-z)$, with a bandwidth $\sigma_m$ that sets how far a window may lie from the line and still be used. With $a_i^K\coloneqq a_i-K(z_i-z)$, the stored action slid along the line to the current history,
\begin{equation}
w_i(c,K)\coloneqq\frac{q^\tau_i(z)\,e^{-\|a_i^K-c\|^2/(2\sigma_m^2)}}{\mu_z(c,K)},\qquad
\mu_z(c,K)\coloneqq\sum_jq^\tau_j(z)\,e^{-\|a_j^K-c\|^2/(2\sigma_m^2)},
\label{eq:mode}
\end{equation}
which are the weights \eqref{eq:cfdp_weights} of the slid actions at $(z,c)$ with bandwidth $\sigma_m$. We call $\mu_z(c,K)$, the fraction of $q^\tau(z)$ on the line, the \emph{support} of the line. 
We refit the line to the weighted windows:
\begin{equation}
(c^{(j+1)},K^{(j+1)})\coloneqq\argmin_{c,K}\sum_iw_i\,\|a_i^K-c\|^2=\big(A\,\Proj_w(z),\,K_w\big),\qquad w=w(c^{(j)},K^{(j)}),
\label{eq:irls}
\end{equation}%
starting from the flat line $(a^\smp,0)$ through a diffusion sample $a^\smp(z,\xi)$. The sample lies close to a stored action $a_i$, chosen approximately with probability $q^\tau_i(z)$ (Section~\ref{sec:policy}), so it provides a starting point near the data. Each refit is the consistent action $A\Proj_w(z)$ of Section~\ref{sec:signed}, with reference weights that down-weight windows off the line. The iteration \eqref{eq:irls} is iteratively reweighted least squares for the robust loss $1-e^{-\|a_i^K-c\|^2/(2\sigma_m^2)}$ \citep{hunter2004tutorial}, so windows off the line receive little weight and do not pull the slope. If the demonstrations contain several distinct ways of acting at the same history (e.g. going left versus going right around an obstacle), the starting point also decides which one the line follows. Note that weighting windows by the distance of their actions to the sample instead would bias the slope toward zero. 

\noindent
{\bf Limiting the extrapolation.}\label{sec:leverage} To avoid trusting the line far from the data, we follow it only up to $r\sqrt{n_\eff(w)}$ standard deviations of the weighted histories, where $n_\eff(w)\coloneqq1/\|w\|^2$ is the effective number of windows:
\begin{equation}
\methodeq{BD-$\chi^2$}{\alpha_w(z)\coloneqq\min\Big\{1,\ \frac{r\sqrt{n_\eff(w)}}{\|S_w^{-1/2}(z-Hw)\|}\Big\},\qquad
u(z)\coloneqq Aw+\alpha_w(z)\,K_w(z-Hw)}
\label{eq:leverage}
\end{equation}

\noindent
{\bf Carrying the line along an episode.}\label{sec:track} Drawing a new diffusion sample at every step changes the starting point and thus the fitted line, which can make the closed-loop behavior erratic. Instead, we carry the line to the next history $z^+$ as $(c+K(z^+-z),K)$, refit it there, and draw a new sample only when its support falls below $\mu_{\min}$, i.e., when the line has left the data.

Putting the three fixes together, one control step of \mthd{BD-$\chi^2$} carries the line from the previous step to the current history $z$ and refits it with $T_{\mathrm{it}}$ iterations of \eqref{eq:irls}, restarts from the flat line through a new diffusion sample only if the support falls below $\mu_{\min}$, and returns the clipped action \eqref{eq:leverage} with the weights of the last refit (Algorithm~\ref{alg:full} in Appendix~\ref{app:alg}).

\vspace{-0.3cm}

\section{Numerical Experiments}\label{sec:experiment}
We perform numerical experiments on two sets of control tasks: the first set consists of the three nonlinear control examples borrowed from \citet{naf2025choose}, and the second set is the state-based Push-T task from \citet{chi2023diffusion} (See Figure~\ref{fig:bd_chi2_demo}).
All experiments are performed with a desktop equipped with an AMD Ryzen 9 7950X CPU.

\begin{table}[!h]
\centering
\caption{Closed-loop results on the three nonlinear control tasks.}
\label{tab:results}
\renewcommand{\arraystretch}{1.15}
\setlength{\tabcolsep}{4pt}
\resizebox{\linewidth}{!}{%
\begin{tabular}{@{}cl*{8}{c}@{}}
\toprule
 & & \mthd{BD-mean} & \mthd{BD-sample} & \mthd{BD-KL} & \mthd{BD-mean-$\chi^2$}
 & \mthd{BD-$\chi^2$} & \mthd{DeePC} & \mthd{Windowed-DPC} & \mthd{Select-DPC} \\
\midrule
\multirow{3}{*}{\rotatebox[origin=c]{90}{Cart-Pole}}
 & Successful Trials~$\uparrow$
 & $0/10$ & $0/10$ & $0/10$ & $10/10$ & $10/10$ & $1/10$ & $0/10$ & $4/10$ \\
 & Cost w.r.t.\ Expert$^{a}$~$\downarrow$
 & --- & --- & --- & $1.194\pm0.108$
 & $1.485\pm0.097$ & $3.903^{b}$ & --- & $3.257\pm1.116$ \\
 & Computation (ms)~$\downarrow$
 & $0.138\pm0.006$ & $0.736\pm0.016$ & ---$^{c}$ & $0.140\pm0.001$ & $0.316\pm0.011$ & $(4.85\pm1.86)\!\times\!10^{3}$ & $124.029\pm7.801$ & $73.041\pm2.884$ \\
\midrule
\multirow{3}{*}{\rotatebox[origin=c]{90}{Rocket}}
 & Successful Trials~$\uparrow$
 & $4/10$ & $3/10$ & $0/10$ & $10/10$ & $10/10$ & $1/10$ & $1/10$ & $6/10$ \\
 & Cost w.r.t.\ Expert$^{a}$~$\downarrow$
 & $1.425\pm0.746$ & $1.404\pm0.636$ & --- & $1.045\pm0.049$
 & $1.036\pm0.047$ & $2.853^{b}$ & $1.050^{b}$ & $1.155\pm0.223$ \\
 & Computation (ms)~$\downarrow$
 & $0.162\pm0.004$ & $1.145\pm0.011$ & ---$^{c}$ & $0.184\pm0.003$ & $0.423\pm0.008$ & $(54.20\pm2.22)\!\times\!10^{3}$ & $(3.32\pm 0.28)\!\times\!10^{3}$ & $(2.50\pm0.60)\!\times\!10^{3}$ \\
\midrule
\multirow{3}{*}{\rotatebox[origin=c]{90}{Reacher}}
 & Successful Trials~$\uparrow$
 & $0/10$ & $0/10$ & $0/10$ & $4/10$ & $7/10$ & $0/10$ & $0/10$ & $7/10$ \\
 & Cost w.r.t.\ Expert$^{a}$~$\downarrow$
 & --- & --- & --- & $1.047\pm0.050$
 & $1.104\pm0.156$ & --- & --- & $1.522\pm0.402$ \\
 & Computation (ms)~$\downarrow$
 & $0.195\pm0.016$ & $1.144\pm0.024$ & ---$^{c}$ & $0.259\pm0.034$ & $0.546\pm0.035$ & $(42.99\pm 20.85)\!\times\!10^{3}$ & $459.310\pm119.924$ & $(2.28\pm1.17)\!\times\!10^{3}$ \\
\bottomrule
\end{tabular}%
}

\vspace{2pt}
\begin{minipage}{\linewidth}
\footnotesize
$^{a}$Mean $\pm$ standard deviation over successful trials only. --- : no successful trial.
$^{b}$Single successful trial, so no standard deviation.
$^{c}$\mthd{BD-KL} returned no actions.
\end{minipage}
\end{table}

\subsection{Control benchmarks}\label{sec:control_benchmark}
{\bf Setup.}
We evaluate all methods in simulation on the three nonlinear control tasks from \citet{naf2025choose}: cart-pole swing-up, rocket landing, and a two-link planar manipulator (i.e., the reacher). For each task, the data are collected by a MPC controller (i.e. the expert), with noise added to its inputs during collection. All methods are tested from the same 10 initial states per task. We report the number of successful trials, the accumulated cost of each successful trial divided by the cost of the expert from the same initial state, and the computation time per control step. The cost is a quadratic penalty on the deviation from the task's reference plus an input penalty. Appendix~\ref{app:tasks} contains the full details of each task.
We compare with three behavioral control baselines: \mthd{DeePC} \citep{coulson2019data}, which uses a fixed data matrix built from all demonstrations, \mthd{Windowed-DPC} \citep{berberich2022linear}, which uses a sliding window of the data, and \mthd{Select-DPC} \citep{naf2025choose}, which selects the data close to the current state at every step. 

\noindent
{\bf Tasks.}
The tasks are cart-pole swing-up, rocket landing, and a two-link reacher that must reach five references in a specified order below a height limit (see details in Appendix~\ref{app:tasks}).



\noindent
{\bf Results.}
Table~\ref{tab:results} reports the results. \mthd{BD-$\chi^2$} succeeds in 27 of the 30 trials. Its three failures on the reacher reach all references in order but exceed the height limit by at most about $1$\,cm. \mthd{Select-DPC} succeeds in 17 of the 30 trials. Its failures come from losing the pole shortly after swinging it up, violating the state constraints of the rocket, or freezing before reaching all references. Among the other methods, \mthd{BD-mean-$\chi^2$} succeeds in 24 trials, while \mthd{BD-mean}, \mthd{BD-sample}, \mthd{BD-KL}, \mthd{DeePC} and \mthd{Windowed-DPC} succeed in at most four trials per task. In terms of computation, \mthd{BD-$\chi^2$} takes $0.3$ to $0.6$\,ms per step, whereas the baselines solve an optimization problem at every step and take from $73$\,ms to $54$\,s.
For further details on the results, see Appendix~\ref{app:control_results}.

\vspace{-0.3cm}
\subsection{Push-T}
{\bf Setup.}
In Push-T \citep{chi2023diffusion}, a circular pusher moves a T-shaped block onto a fixed target pose in the plane. We use the state-based version with all 206 human demonstrations released by \citet{chi2023diffusion}.
The observation $y$ contains the positions of the pusher and the block and the cosine and sine of the block angle. The input $u$ is a target position for the pusher, which is tracked by a PD controller.
As in \citet{chi2023diffusion}, the score of an episode is the largest fraction of the target area covered by the block, divided by $0.95$ and capped at $1$, and a trial is successful once the coverage exceeds $95\%$ within 300 steps.
Both methods are tuned on 8 validation initial states and tested from 50 further initial states drawn by the simulator. Appendix~\ref{app:pusht} gives the details.

\begin{figure}[t]
\centering
\begin{minipage}[c]{0.57\textwidth}
  \centering
  \includegraphics[width=\linewidth]{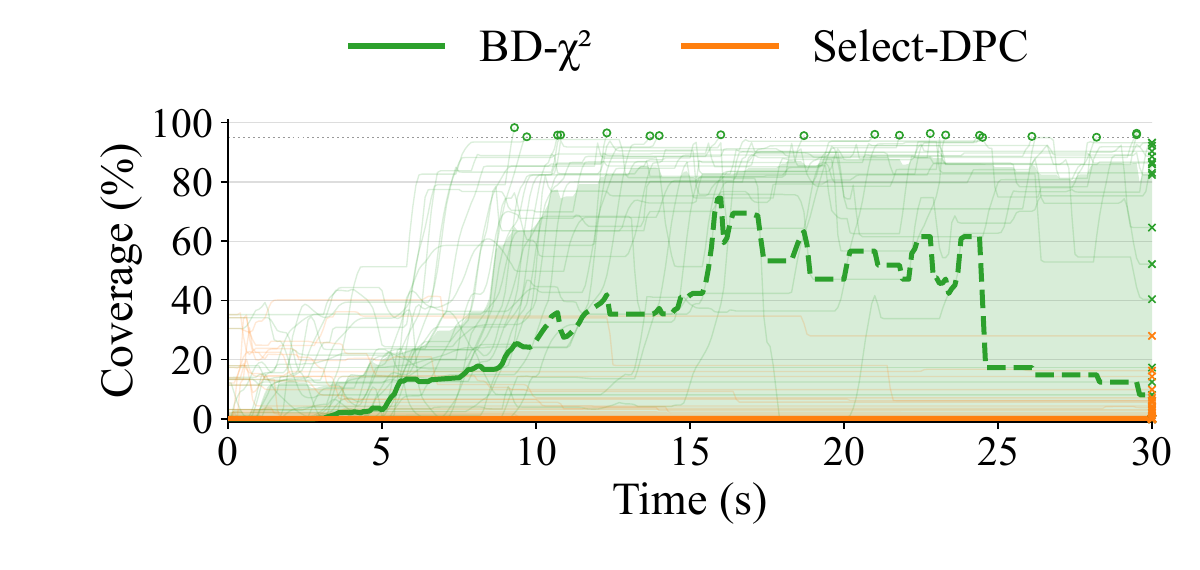}
\end{minipage}\hfill
\begin{minipage}[c]{0.40\textwidth}
  \centering
  \renewcommand{\arraystretch}{1.15}
  \setlength{\tabcolsep}{2pt}
  \small
  \begin{tabular}{@{}lcc@{}}
  \toprule
   & \mthd{BD-$\chi^2$} & \mthd{Select-DPC} \\
  \midrule
  Success~$\uparrow$ & $19/50$ & $0/50$ \\
  Score~$\uparrow$ & $0.66\pm0.45$ & $0.07\pm0.12$ \\
  Median~$\uparrow$ & $0.98$ & $0.00$ \\
  \bottomrule
  \end{tabular}
\end{minipage}
\vspace{-0.5cm}
\caption{Push-T results over 50 trials. {\bf Left:} Target coverage over time for \mthd{BD-$\chi^2$} (green) and \mthd{Select-DPC} (orange). Thin curves show individual trials, and thick curves and bands show the median and interquartile range over the running trials (dashed once some have ended). $\circ$ marks a success and $\times$ a timeout. The dotted line marks the 95\% success threshold. {\bf Right:} Number of successful trials, and the mean $\pm$ standard deviation and the median of the score over all 50 trials.}
\label{fig:pusht}
\vspace{-0.8cm}
\end{figure}




\noindent
{\bf Results.}
Figure~\ref{fig:pusht} reports the results. \mthd{BD-$\chi^2$} succeeds from 19 of the 50 initial states and comes close on 13 others, where it reaches a score of at least $0.9$ without crossing the success threshold. From 10 initial states, the block never overlaps the target under either method.
\mthd{Select-DPC} succeeds from none: its highest score is $0.44$, and its median score is $0$. Visualizations of closed-loop trajectories can be found in Figure~\ref{fig:pusht_trajectory_comparison} in the appendix. 

\vspace{-0.4cm}

\section{Conclusions}\label{sec:conclusion}
In this work, we took inspirations from closed-form diffusion policy and identified a potential connection to the behavioral control framework. The derivations that yield this connection naturally lead to the development of several algorithm variants, which we tested on several nonlinear control examples.
We hope the presented framework offers a different perspective for approaching the development of nonlinear behavioral control algorithms and helps shed light on understanding deep-learning-based control approaches such as diffusion policies.

A limitation of the proposed algorithm is that, unlike the \mthd{DeePC} family, it does not explicitly enforce constraints or include additional cost terms. 

\vspace{-0.4cm}
\subsection*{Generative AI Use}
Generative AI tools were used for coding, improving the writing, and assisting with proofs.




\bibliography{references}

\appendix
\clearpage
\section{Interpretation of $K_w$}\label{app:Kw}

Since $W_w=\mathrm{diag}(w)-ww^\top$, the matrices in \eqref{eq:moments} are the weighted sums
\[
S_w=\sum_iw_i(z_i-Hw)(z_i-Hw)^\top,\qquad
AW_wH^\top=\sum_iw_i(a_i-Aw)(z_i-Hw)^\top.
\]%

$K_w$ in \eqref{eq:moments} is the slope of the $w$-weighted least-squares fit of the actions on the histories. In the scalar case it is $\Cov_w(a,z)/\mathrm{Var}_w(z)$. In general, entry $(j,k)$ is the sensitivity of action coordinate $j$ to history coordinate $k$ with the other history coordinates held fixed, which is what the factor $S_w^{-1}$ accounts for.

$K_w$ can also be interpreted as an estimate of the feedback gain of the demonstrator (i.e. the expert). If the actions were generated by an affine feedback law, $a_i=Gz_i+c_0$ for all $i$, then $a_i-Aw=G(z_i-Hw)$, and $K_w=G$ for every $w$ with $S_w\succ0$. If they were generated by a smooth feedback law, $a_i=\kappa(z_i)$, and $w$ puts its weight on windows whose histories are close to $Hw$, then $K_w\approx\nabla\kappa(Hw)$.

\section{Different Policies under a Unifying Framework}\label{app:template}
\begin{table}[h]
\centering
\caption{The policies discussed in this paper as instances of \eqref{eq:template}.}
\label{tab:template}
\small
\begin{tabular}{@{}lllll@{}}
\toprule
Policy & Reference $w$ & $D_f$ & $\lambda$ & $\zeta$ \\
\midrule
Behavioral Control \eqref{eq:bc} & uniform & $\chi^2$ & $>0$ & $z$ \\
\mthd{BD-mean} \eqref{eq:mean} & $q^\tau(z)$ & -- & $\infty$ & -- \\
\mthd{BD-sample} \eqref{eq:ode} & $g^\sigma(z,\tilde a)$, $\sigma$ annealed & -- & $\infty$ & -- \\
\mthd{BD-KL} \eqref{eq:klproj} & $g^\sigma(z,\tilde a)$ & $\KL$ & $0$ & $z$ \\
\mthd{BD-mean-$\chi^2$} \eqref{eq:meanchi2} & $q^\tau(z)$ & $\chi^2$ & $0$ & $z$ \\
\mthd{BD-$\chi^2$} \eqref{eq:leverage} & $w(c,K)$ \eqref{eq:mode} & $\chi^2$ & $0$ & $Hw+\alpha_w(z)(z-Hw)$ \\
\bottomrule
\end{tabular}
\end{table}

\section{Proof of the Spread Identity}\label{app:split}

\noindent\textbf{Identity \eqref{eq:split}.} \emph{Let $M$ have columns $m_1,\dots,m_N$ and let $V^M(g)\coloneqq\sum_ig_i\|m_i-Mg\|^2$. For $g\in\Delta_N$ and $x$ of appropriate dimension,}
\[
\|Mg-x\|^2+V^M(g)=\sum_{i=1}^Ng_i\|m_i-x\|^2.
\tag{\ref{eq:split}}
\]

\begin{proof}
Let $\bar m\coloneqq Mg$. For each $i$,
\[
\|m_i-x\|^2=\|m_i-\bar m\|^2+2(m_i-\bar m)^\top(\bar m-x)+\|\bar m-x\|^2.
\]
Multiply by $g_i$ and sum over $i$. The first terms give $V^M(g)$. The cross terms vanish, because $\sum_ig_i(m_i-\bar m)=Mg-\bar m\sum_ig_i=\bar m-\bar m=0$. The last terms give $\|\bar m-x\|^2\sum_ig_i=\|Mg-x\|^2$.
\end{proof}

\section{Exact and Discretized Sampling in \mthd{BD-sample}}\label{app:flow_endpoint}
For the exact flow and $\sigma_{\max}\to\infty$, the endpoint is a stored action $a_i$ with probability $q^\tau_i(z)$. With finitely many Euler steps, the last step, from the smallest nonzero noise level $\sigma_{\min}$ to $0$, returns $D(z,\tilde a;\sigma_{\min})$: a convex combination of the stored actions that concentrates on one of them when $\sigma_{\min}$ is small.

\section{Proof of Proposition~\ref{prop:approx}}\label{app:prop_approx}

\noindent\textbf{Proposition~\ref{prop:approx}.} \emph{Let $\kappa:\mathcal Z\to\R^{n_a}$ be differentiable on an open convex set $\mathcal Z$ containing $z$ and the $z_i$, with $\|\nabla\kappa\|\le L$ and $\nabla\kappa$ $B$-Lipschitz on $\mathcal Z$, and let $\eta_i\coloneqq a_i-\kappa(z_i)$ be the deviation of window $i$'s action from $\kappa$. For every $\zeta\in\mathcal Z$ and $g\in\Cset(\zeta)$,}
\[
\Big\|Ag-\kappa(z)\Big\|\le L\,\|z-\zeta\|+\frac B2\sum_i|g_i|\,\|z_i-\zeta\|^2+\Big\|\sum_ig_i\eta_i\Big\|.
\]

\begin{proof}
By the definition of $\eta_i$, $a_i=\kappa(z_i)+\eta_i$ for each $i$. Let
\[
R_i\coloneqq\kappa(z_i)-\kappa(\zeta)-\nabla\kappa(\zeta)(z_i-\zeta)
\]
be the error of the first-order approximation of $\kappa(z_i)$ around $\zeta$. Combining the two gives, for each $i$,
\begin{equation}
a_i=\kappa(\zeta)+\nabla\kappa(\zeta)(z_i-\zeta)+R_i+\eta_i.
\label{eq:aiexpand}
\end{equation}
The segment from $\zeta$ to $z_i$ lies in $\mathcal Z$ because $\mathcal Z$ is convex, so by the fundamental theorem of calculus,
\[
R_i=\int_0^1\Big(\nabla\kappa\big(\zeta+s(z_i-\zeta)\big)-\nabla\kappa(\zeta)\Big)(z_i-\zeta)\,ds.
\]
Since $\nabla\kappa$ is $B$-Lipschitz, the difference of Jacobians in the integrand has norm at most $Bs\|z_i-\zeta\|$, and therefore
\begin{equation}
\|R_i\|\le\int_0^1Bs\,\|z_i-\zeta\|^2\,ds=\tfrac B2\|z_i-\zeta\|^2.
\label{eq:Ribound}
\end{equation}

Next, we multiply \eqref{eq:aiexpand} by $g_i$ and sum over $i$. The left-hand side becomes $\sum_ig_ia_i=Ag$. On the right-hand side, the first term becomes $\kappa(\zeta)\sum_ig_i=\kappa(\zeta)$, because $\sum_ig_i=1$. The second term becomes
\[
\nabla\kappa(\zeta)\Big(\sum_ig_iz_i-\zeta\sum_ig_i\Big)=\nabla\kappa(\zeta)\,(Hg-\zeta)=0,
\]
because $Hg=\zeta$ for $g\in\Cset(\zeta)$. Hence $Ag=\kappa(\zeta)+\sum_ig_iR_i+\sum_ig_i\eta_i$, and subtracting $\kappa(z)$ from both sides gives
\begin{equation}
Ag-\kappa(z)=\big(\kappa(\zeta)-\kappa(z)\big)+\sum_ig_iR_i+\sum_ig_i\eta_i.
\label{eq:errsplit}
\end{equation}

Finally, we take norms in \eqref{eq:errsplit}. By the triangle inequality,
\[
\|Ag-\kappa(z)\|\le\|\kappa(\zeta)-\kappa(z)\|+\Big\|\sum_ig_iR_i\Big\|+\Big\|\sum_ig_i\eta_i\Big\|,
\]
and we bound the first two terms on the right. For the first, the segment from $z$ to $\zeta$ lies in $\mathcal Z$, so by the fundamental theorem of calculus and $\|\nabla\kappa\|\le L$,
\[
\|\kappa(\zeta)-\kappa(z)\|=\Big\|\int_0^1\nabla\kappa\big(z+s(\zeta-z)\big)(\zeta-z)\,ds\Big\|\le L\,\|\zeta-z\|.
\]
For the second, the triangle inequality and \eqref{eq:Ribound} give
\[
\Big\|\sum_ig_iR_i\Big\|\le\sum_i|g_i|\,\|R_i\|\le\frac B2\sum_i|g_i|\,\|z_i-\zeta\|^2,
\]
where the absolute values are needed because the weights in $\Cset(\zeta)$ may be negative. Inserting these two bounds gives the claim.
\end{proof}

The proof splits the error of the action, $Ag-\kappa(z)$, into the three terms on the right-hand side of \eqref{eq:errsplit}, and each term of the bound in Proposition~\ref{prop:approx} bounds one of them. The three terms have different causes and respond differently to the choice of the weights $g$, so we discuss them one by one.

\emph{First-order error, $\kappa(\zeta)-\kappa(z)$.} The weights combine the stored histories into $\zeta=Hg$. If $\zeta$ differs from the current history $z$, the combination answers the question of what the expert would do at $\zeta$, not at $z$. The error this causes is at most $L\|z-\zeta\|$, where $L$ bounds how quickly the expert's action changes with the history. It vanishes for consistent weights ($\zeta=z$), which is the reason for adding the constraint $Hg=z$ in Section~\ref{sec:constraint}.

\emph{Curvature error, $\sum_ig_iR_i$.} Each stored action $a_i$ is the expert's response at its own history $z_i$, not at $\zeta$. Whether combining such actions gives the response at $\zeta$ depends on the shape of $\kappa$. If $\kappa$ is affine, it does: the weights average the histories $z_i$ to exactly $\zeta$, and an affine function of an average is the average of its values, so the combined action is exactly $\kappa(\zeta)$. If $\kappa$ is curved, the average of its values is no longer its value at the average, and the difference is $\sum_ig_iR_i$.

For example, let $\kappa(z)=z^2$ and take two windows at $z_1=d$ and $z_2=-d$ with equal weights $g_1=g_2=\frac12$. The combined history is $\zeta=0$, where the expert's action is $\kappa(0)=0$. The combined action, however, is $\frac12d^2+\frac12d^2=d^2$. The error $d^2$ equals the bound of Proposition~\ref{prop:approx}, since $B=2$ for this $\kappa$ and $\frac B2\sum_i|g_i|\,\|z_i-\zeta\|^2=d^2$.

The contribution of window $i$ is at most $\frac B2|g_i|\,\|z_i-\zeta\|^2$. It grows with the square of the distance between $z_i$ and $\zeta$, so a window twice as far away can contribute four times as much. The curvature error is therefore small when the weight sits on windows whose histories are close to $\zeta$, and large when distant windows carry weight. For $g\in\Delta_N$ and $\zeta=Hg$, the bound is $\frac B2\sum_ig_i\|z_i-Hg\|^2=\frac B2V^H(g)$. This is the spread of the histories that the spread term of \eqref{eq:dbc} penalizes, which is why behavioral denoising keeps the curvature error small. Finally, the bound contains $|g_i|$, so a negative weight contributes as much as a positive weight of the same size. Since the weights sum to one, using negative weights means that $\sum_i|g_i|>1$, and the bound is then larger than for nonnegative weights on the same windows.

\emph{Noise error, $\sum_ig_i\eta_i$.} The deviation $\eta_i=a_i-\kappa(z_i)$ is the part of the stored action $a_i$ that the reference law $\kappa$ does not explain. The combined action inherits these deviations with the same weights as the actions. The deviations come from two sources, which behave differently.
The first source is randomness, such as measurement noise and small variations in how the expert acts.
Suppose that these deviations are independent of each other and of the weights, with zero mean and $\E\|\eta_i\|^2=s^2$. Then $\E\|\sum_ig_i\eta_i\|^2=s^2\sum_ig_i^2$, because the cross terms have zero mean. The factor $\sum_ig_i^2$ measures how concentrated the weights are. If the weight is spread evenly over $n$ windows, it equals $1/n$, and the noise is averaged down by a factor of $n$. If one window carries all the weight, it equals $1$, and the combined action inherits the full noise of that window. In the notation of Section~\ref{sec:fixes}, $\sum_ig_i^2=1/n_\eff(g)$. With negative weights, $\sum_ig_i^2$ can exceed $1$, in which case the combination amplifies the noise instead of averaging it. Note that this pulls in the opposite direction to the curvature error: averaging the noise calls for spreading the weight over many windows, while a small curvature error calls for keeping it on the few windows closest to $\zeta$.
The second source is the expert acting in more than one way. When the expert responds to similar histories in several distinct ways, for example by passing an obstacle sometimes on the left and sometimes on the right, $\kappa$ describes one of them, and every window of the other way has a large deviation $\eta_i$ in the same direction. These deviations do not have zero mean, so they do not cancel however many windows are averaged. Instead, they pull the combined action toward the other way of acting by an amount proportional to the weight on its windows. The result lies between the two ways of acting and belongs to neither, such as steering straight at the obstacle.

\section{The Spread Term}\label{app:spread}\label{app:prop_beta}

This section expands Section~\ref{sec:spread}. To see the effect of the spread term, scale it by $\beta\in[0,1]$:
\begin{equation}
\min_{g\in\Delta_N}\ \frac{\|Hg-z\|^2+\beta V^H(g)}{2\tau^2}+\frac{\|Ag-\tilde a\|^2+\beta V^A(g)}{2\sigma^2}+\KL(g\,\|\,\pi).
\label{eq:beta}
\end{equation}
At $\beta=1$ this is behavioral denoising \eqref{eq:dbc}. At $\beta=0$ it is the reconstruction problem \eqref{eq:bc} with an action term added, restricted to the simplex, and with KL as the regularizer. With $\phi_i=\phi_i(z,\tilde a)$ of \eqref{eq:cfdp_weights}, and apply \eqref{eq:split} to $H$ and to $A$, the first two terms become:
\begin{equation}
(1-\beta)\Big(\frac{\|Hg-z\|^2}{2\tau^2}+\frac{\|Ag-\tilde a\|^2}{2\sigma^2}\Big)+\beta\sum_ig_i\phi_i.
\label{eq:betasplit}
\end{equation}
The coefficient $\beta$ therefore balances two objectives: \emph{reconstruction} asks the combination $(Hg,Ag)$ to reproduce $(z,\tilde a)$, while \emph{locality} asks every window that is used to have its history close to $z$ and its action close to $\tilde a$.

\begin{proposition}\label{prop:beta}
Let $\tau,\sigma>0$ and $\pi\in\Delta_N^\circ$. For $\beta\in[0,1]$, \eqref{eq:beta} has a unique minimizer $g^\beta\in\Delta_N^\circ$, and
\[
\begin{gathered}
g^\beta_i\propto\pi_i\exp\!\Big(-\beta\phi_i+\theta_z^\top(z_i-z)+\theta_a^\top(a_i-\tilde a)\Big),\\
\theta_z\coloneqq-\frac{1-\beta}{\tau^2}(Hg^\beta-z),\qquad\theta_a\coloneqq-\frac{1-\beta}{\sigma^2}(Ag^\beta-\tilde a).
\end{gathered}
\]
\end{proposition}


\begin{proof}
By \eqref{eq:betasplit}, on $\Delta_N$ the objective of \eqref{eq:beta} is
\[
F(g)\coloneqq(1-\beta)\Big(\frac{\|Hg-z\|^2}{2\tau^2}+\frac{\|Ag-\tilde a\|^2}{2\sigma^2}\Big)+\beta\sum_ig_i\phi_i+\sum_ig_i\log\frac{g_i}{\pi_i},
\]
which is a convex quadratic in $g$, plus a linear term, plus the strictly convex entropy term $\sum_ig_i\log(g_i/\pi_i)$.

\emph{Existence and uniqueness.} $F$ is continuous and strictly convex on the compact set $\Delta_N$, so it has a unique minimizer $g^\beta$.

\emph{The minimizer lies in $\Delta_N^\circ$.} Suppose instead that $g^\beta_i=0$ for some $i$, and pick $j$ with $g^\beta_j>0$. Moving a small mass $\varepsilon>0$ from window $j$ to window $i$ keeps the weights in $\Delta_N$. It changes the entropy term of window $i$ from $0$ to $\varepsilon\log(\varepsilon/\pi_i)=\varepsilon\log\varepsilon+O(\varepsilon)$, and it changes the entropy term of window $j$ and the quadratic and linear terms by $O(\varepsilon)$, because these are differentiable at $g^\beta$. The objective therefore changes by:
\[
\varepsilon\log\varepsilon+O(\varepsilon)=\varepsilon\,\big(\log\varepsilon+O(1)\big),
\]
which is negative for small $\varepsilon$, contradicting the optimality of $g^\beta$.

\emph{Stationarity.} Since $g^\beta\in\Delta_N^\circ$, the constraint $g\ge0$ is inactive and $F$ is differentiable at $g^\beta$. The only remaining constraint is $\one^\top g=1$, so the derivative of $F$ at $g^\beta$ must vanish along every direction $d$ with $\one^\top d=0$. This means that all partial derivatives $\partial F/\partial g_i$ at $g^\beta$ are equal to a common value $\nu$, the Lagrange multiplier of the constraint:
\[
(1-\beta)\Big(\frac{z_i^\top(Hg^\beta-z)}{\tau^2}+\frac{a_i^\top(Ag^\beta-\tilde a)}{\sigma^2}\Big)+\beta\phi_i+\log\frac{g^\beta_i}{\pi_i}+1=\nu\quad \forall i.
\]
By the definitions of $\theta_z$ and $\theta_a$, the first term equals $-\theta_z^\top z_i-\theta_a^\top a_i$, hence:
\[
\log\frac{g^\beta_i}{\pi_i}=\nu-1-\beta\phi_i+\theta_z^\top z_i+\theta_a^\top a_i.
\]

\emph{The stated form.} Write $z_i=(z_i-z)+z$ and $a_i=(a_i-\tilde a)+\tilde a$, so that $\theta_z^\top z_i=\theta_z^\top(z_i-z)+\theta_z^\top z$ and $\theta_a^\top a_i=\theta_a^\top(a_i-\tilde a)+\theta_a^\top\tilde a$. The terms $\theta_z^\top z$ and $\theta_a^\top\tilde a$ are the same for every window, and so is $\nu-1$. Collecting them in $C\coloneqq\nu-1+\theta_z^\top z+\theta_a^\top\tilde a$ and exponentiating gives
\[
g^\beta_i=e^{C}\,\pi_i\exp\big(-\beta\phi_i+\theta_z^\top(z_i-z)+\theta_a^\top(a_i-\tilde a)\big).
\]
The factor $e^{C}$ does not depend on $i$ and is fixed by $\sum_ig^\beta_i=1$, which is the stated form. For $\beta<1$, $\theta_z$ and $\theta_a$ are themselves defined through $g^\beta$, so this form is an equation that $g^\beta$ satisfies rather than a formula for computing it. For $\beta=1$, $\theta_z=\theta_a=0$ and the form is explicit.
\end{proof}

The two ends behave in opposite ways.
\begin{itemize}
\item \emph{$\beta=0$: reconstructing, not local.} The weights are the prior reweighted by $e^{\theta_z^\top z_i+\theta_a^\top a_i}$ (up to normalization). Setting $\beta=0$ in the definitions of $\theta_z$ and $\theta_a$ in Proposition~\ref{prop:beta} and rearranging gives
\[
Hg^0-z=-\tau^2\theta_z,\qquad Ag^0-\tilde a=-\sigma^2\theta_a.
\]
The reweighting direction $(\theta_z,\theta_a)$ (i.e. the tilt) and the reconstruction error are therefore the same up to scaling: $\theta_z$ points from the combined history $Hg^0$ toward $z$, and $\theta_a$ from the combined action $Ag^0$ toward $\tilde a$. 
The factor $e^{\theta_z^\top z_i+\theta_a^\top a_i}$ thus puts more weight on windows that lie further in the direction in which the combination still falls short of $(z,\tilde a)$, which moves the combination toward $(z,\tilde a)$. 
Reweighting more strongly reduces the reconstruction error but moves the weights further from the prior $\pi$, which the KL term penalizes. The minimizer $g^0$ is the point at which reducing the error further would cost more in KL divergence than it saves in the first two terms of \eqref{eq:beta}, and the two identities above state how much error is left at that point: $\tau^2$ times $\theta_z$ for the history and $\sigma^2$ times $\theta_a$ for the action. The remaining error is therefore small when $\tau$ and $\sigma$ are small, which makes errors expensive, or when a mild reweighting already suffices.
However, the tilt selects windows by how far they lie along the direction $(\theta_z,\theta_a)$, not by how close they are to $(z,\tilde a)$, so the windows it mixes may come from anywhere in the data set, including parts unrelated to the current history.
\item \emph{$\beta=1$: local, not reconstructing.} Here $\theta_z=\theta_a=0$, and the weights are the behavioral denoiser $g^\sigma(z,\tilde a)$. Each weight decays with $\phi_i$, so the combination uses only windows whose history is close to $z$ and whose action is close to $\tilde a$. What is lost is reconstruction: by \eqref{eq:histresidual}, the combined history misses $z$ by $Hg^\sigma-z=\tau^2\nabla_z\log\tilde p(z,\tilde a;\tau,\sigma)$. In words, the combined history is a denoised version of $z$, pulled toward where nearby windows are dense.
\end{itemize}
In terms of Proposition~\ref{prop:approx}, $\beta=1$ keeps the curvature error small but leaves a first-order error, and $\beta=0$ reverses the trade-off.

\section{Behavioral Denoising as a Policy}\label{app:limits}\label{app:cor_mean}

This section expands Section~\ref{sec:limits}.

\emph{The reference law $\kappa$.} The feedback law $\kappa$ in Proposition~\ref{prop:approx} describes how the expert responds to the history, and serves as the reference against which a policy is compared. The recorded actions need not equal $\kappa(z_i)$: for example, when the expert sometimes passes an obstacle on the left and sometimes on the right, and $\kappa$ describes passing on the left, the windows that pass on the right have actions far from $\kappa(z_i)$, i.e., large deviations $\eta_i$.

\emph{The mean policy.} For \mthd{BD-mean} \eqref{eq:mean}, the weights $q^\tau(z)$ lie in $\Delta_N$, and Proposition~\ref{prop:approx} takes the following form.

\begin{corollary}\label{cor:mean}
Under the assumptions of Proposition~\ref{prop:approx}, the \mthd{BD-mean} policy \eqref{eq:mean} satisfies
\[
\|u_{\mathrm{mean}}(z)-\kappa(z)\|\le L\,\tau^2\|\nabla_z\log\tilde p(z;\tau)\|+\frac B2V^H(q^\tau(z))+\Big\|\sum_iq^\tau_i(z)\eta_i\Big\|,
\]
where $\tilde p(z;\tau)\coloneqq\sum_j\pi_j\N(z;z_j,\tau^2\I)$ is the density of the stored histories after adding Gaussian noise of standard deviation $\tau$.
\end{corollary}

\begin{proof}
Throughout the proof, write $q\coloneqq q^\tau(z)$ for the weights of \mthd{BD-mean} and $\zeta\coloneqq Hq$ for the history they combine to. We apply Proposition~\ref{prop:approx} with $g=q$ and this $\zeta$, and then rewrite the three terms of its bound.

Since $Aq=u_{\mathrm{mean}}(z)$ by \eqref{eq:mean}, Proposition~\ref{prop:approx} gives
\[
\|u_{\mathrm{mean}}(z)-\kappa(z)\|\le L\,\|z-Hq\|+\frac B2\sum_i|q_i|\,\|z_i-Hq\|^2+\Big\|\sum_iq_i\eta_i\Big\|.
\]
The third term is already the third term of the claim.

\emph{The second term.} Since $q_i\ge0$, we have $|q_i|=q_i$, and the sum is $\sum_iq_i\|z_i-Hq\|^2=V^H(q)$ by the definition of the spread $V^H$ in Section~\ref{sec:objective}. The second term is therefore $\frac B2V^H(q^\tau(z))$.

\emph{The first term.} It remains to show that
\begin{equation}
Hq-z=\tau^2\,\nabla_z\log\tilde p(z;\tau).
\label{eq:meanscore}
\end{equation}
The gradient of a Gaussian density with respect to its argument is $\nabla_z\N(z;z_j,\tau^2\I)=\N(z;z_j,\tau^2\I)\,(z_j-z)/\tau^2$. Differentiating $\tilde p(z;\tau)=\sum_j\pi_j\N(z;z_j,\tau^2\I)$ term by term and dividing by $\tilde p(z;\tau)$ gives
\[
\nabla_z\log\tilde p(z;\tau)=\frac{\nabla_z\tilde p(z;\tau)}{\tilde p(z;\tau)}=\frac1{\tau^2}\sum_j\frac{\pi_j\,\N(z;z_j,\tau^2\I)}{\tilde p(z;\tau)}\,(z_j-z).
\]
The fraction inside the sum is $q_j$: it is proportional to $\pi_je^{-\|z_j-z\|^2/(2\tau^2)}$, as in the definition of $q^\tau_j(z)$, and it sums to one over $j$ because $\tilde p(z;\tau)$ is the sum of the numerators. Therefore
\[
\nabla_z\log\tilde p(z;\tau)=\frac1{\tau^2}\sum_jq_j\,(z_j-z)=\frac1{\tau^2}\,(Hq-z),
\]
where the last step uses $\sum_jq_j=1$. This shows \eqref{eq:meanscore}, which is the counterpart of the first identity of \eqref{eq:histresidual} for the histories alone.
\end{proof}

\section{Proofs for Section~\ref{sec:consistency}}\label{app:sec3}



\begin{proposition}\label{prop:klproj}\label{app:klproj}
Let $\tau,\sigma>0$, $\pi\in\Delta_N^\circ$ and $z\in\conv\{z_1,\dots,z_N\}$. The minimizer $g^{\mathrm c}(z,\tilde a;\sigma)$ of \eqref{eq:dbc} over $\Delta_N\cap\Cset(z)$ exists, is unique, and equals \eqref{eq:klproj}, the KL projection of the behavioral denoiser onto consistent weights. If $z\in\ri\conv\{z_1,\dots,z_N\}$, then $g^{\mathrm c}\in\Delta_N^\circ$ and $g^{\mathrm c}_i\propto g^\sigma_i(z,\tilde a)\,e^{\theta^\top(z_i-z)}$, where $\theta$ minimizes $\psi(\theta)\coloneqq\log\sum_ig^\sigma_i(z,\tilde a)e^{\theta^\top(z_i-z)}$.
\end{proposition}

\begin{remark}
    As $\sigma\to\infty$, $g^{\mathrm c}$ becomes the KL projection of $q^\tau(z)$ onto the consistent weights, the local maximum-entropy approximation of \citet{arroyo2006local}.
\end{remark}

\begin{proof}
Write $g^\sigma$ for $g^\sigma(z,\tilde a)$.
By \eqref{eq:dbc_kl}, on $\Delta_N$ the objective of \eqref{eq:dbc} equals $\KL(g\,\|\,g^\sigma)$ up to a constant that does not depend on $g$. Minimizing \eqref{eq:dbc} over $\Delta_N\cap\Cset(z)$ is therefore the same as minimizing $\KL(g\,\|\,g^\sigma)$ over this set, which is \eqref{eq:klproj}.
The feasible set $\Delta_N\cap\Cset(z)$ consists of the weights with $g\ge0$, $\one^\top g=1$, and $Hg=z$. It is nonempty, because $z\in\conv\{z_i\}$ means that $z=Hg$ for some $g\in\Delta_N$. It is also convex and compact. Since $\KL(\cdot\,\|\,g^\sigma)$ is continuous and strictly convex on it, the minimizer $g^{\mathrm c}$ exists and is unique.

For the rest of the proof, let $z\in\ri\conv\{z_i\}$.

\emph{The minimizer lies in $\Delta_N^\circ$.} Since $z$ lies in the relative interior of the convex hull, it can be written as $z=H\bar g$ with weights $\bar g\in\Delta_N^\circ$, so the feasible set contains a strictly positive point $\bar g$. The claim now follows as in the proof of Proposition~\ref{prop:beta}, with one change: instead of moving a mass $\varepsilon$ between two windows, which would violate $Hg=z$, we move from $g^{\mathrm c}$ a small step $\varepsilon$ toward $\bar g$, which stays in the feasible set. If some entries of $g^{\mathrm c}$ were zero, this step would make them positive and change the objective by a negative multiple of $\varepsilon\log(1/\varepsilon)$ plus $O(\varepsilon)$, which is negative for small $\varepsilon$ and contradicts the optimality of $g^{\mathrm c}$.

\emph{Stationarity.} Since $g^{\mathrm c}\in\Delta_N^\circ$, the constraint $g\ge0$ is inactive, and $\KL(\cdot\,\|\,g^\sigma)$ is differentiable at $g^{\mathrm c}$, with partial derivatives $\log(g_i/g^\sigma_i)+1$. The remaining constraints, $\one^\top g=1$ and $Hg=z$, are linear. At the minimizer, the objective cannot decrease along any direction $d$ that keeps both of them satisfied, i.e., any $d$ with $\one^\top d=0$ and $Hd=0$. The gradient of the objective at $g^{\mathrm c}$ is therefore orthogonal to all such directions, which means that it is a linear combination of the gradients of the constraints: there are multipliers $\nu\in\R$ for $\one^\top g=1$ and $\theta\in\R^{n_z}$ for $Hg=z$ such that
\[
\log\frac{g^{\mathrm c}_i}{g^\sigma_i}+1=\nu+\theta^\top z_i\quad \forall i.
\]

\emph{The stated form.} Solving for $g^{\mathrm c}_i$ and writing $z_i=(z_i-z)+z$ gives
\[
g^{\mathrm c}_i=e^{\nu-1+\theta^\top z}\,g^\sigma_i\,e^{\theta^\top(z_i-z)}.
\]
The first factor is the same for every window and is fixed by $\sum_ig^{\mathrm c}_i=1$: it equals $1/\sum_jg^\sigma_je^{\theta^\top(z_j-z)}=e^{-\psi(\theta)}$. Hence $g^{\mathrm c}=g(\theta)$, where
\[
g_i(\theta)\coloneqq g^\sigma_i\,e^{\theta^\top(z_i-z)-\psi(\theta)},
\]
which is the stated form.

\emph{$\theta$ minimizes $\psi$.} For any $\theta$, the weights $g(\theta)$ are positive and sum to one, but they need not be consistent. Differentiating $\psi$ shows how far they are from it:
\begin{equation}
\nabla\psi(\theta)=\frac{\sum_ig^\sigma_ie^{\theta^\top(z_i-z)}(z_i-z)}{\sum_jg^\sigma_je^{\theta^\top(z_j-z)}}=\sum_ig_i(\theta)(z_i-z)=Hg(\theta)-z.
\label{eq:psiderivs}
\end{equation}
So $g(\theta)$ is consistent exactly when $\nabla\psi(\theta)=0$. The function $\psi$ is a log-sum-exp of affine functions of $\theta$ and hence convex, so its gradient vanishes exactly at its minimizers. For the multiplier $\theta$ found above, $g(\theta)=g^{\mathrm c}$ is consistent, so $\nabla\psi(\theta)=0$ and $\theta$ minimizes $\psi$.

\emph{Every minimizer of $\psi$ gives $g^{\mathrm c}$.} Conversely, let $\theta'$ be any minimizer of $\psi$. Then $\nabla\psi(\theta')=0$, so $g(\theta')$ is consistent, and it satisfies the stationarity condition above with $\theta'$ in place of $\theta$. Because the objective is convex and the constraints are linear, a feasible point that satisfies the stationarity condition is a minimizer. By uniqueness, $g(\theta')=g^{\mathrm c}$. In particular, $g^{\mathrm c}$ can be computed by minimizing $\psi$, and it does not matter which minimizer is found.
\end{proof}



\subsection{Derivation of \eqref{eq:chi2}}\label{app:chi2}

\noindent\textbf{Claim \eqref{eq:chi2}.} \emph{For $w\in\Delta_N^\circ$ and $g$ near $w$ with $\one^\top g=1$,}
\[
\KL(g\,\|\,w)=\tfrac12\chi^2(g\,\|\,w)+O\big(\|g-w\|^3\big), \text{ where } \chi^2(g\,\|\,w)\coloneqq\sum_i\frac{(g_i-w_i)^2}{w_i}.
\]

\begin{proof}
Let $\delta\coloneqq g-w$, so $\one^\top\delta=0$. The expansion $\log(1+x)=x-x^2/2+O(|x|^3)$, applied with $x=\delta_i/w_i$, gives for each $i$
\[
g_i\log\frac{g_i}{w_i}=(w_i+\delta_i)\log\Big(1+\frac{\delta_i}{w_i}\Big)=\delta_i+\frac{\delta_i^2}{2w_i}+O\Big(\frac{|\delta_i|^3}{w_i^2}\Big).
\]
The remainder carries the factor $1/w_i^2$ because the expansion is in the ratio $\delta_i/w_i$. Bounding this factor by $1/(\min_jw_j)^2$ makes the remainder $O(|\delta_i|^3)$ for all $i$ at once, with a constant that depends on $\min_jw_j$. Summing over $i$, $\sum_i\delta_i=0$, which leaves $\frac12\chi^2(g\|w)+O(\|\delta\|^3)$.
\end{proof}


\subsection{Proof of Proposition~\ref{prop:proj}}\label{app:proj}

\noindent\textbf{Proposition~\ref{prop:proj}.} \emph{Let $w\in\Delta_N^\circ$ with $S_w\succ0$. Then $\Proj_w(\zeta)$ is unique,}
\[
\Proj_w(\zeta)=w+W_wH^\top S_w^{-1}(\zeta-Hw),
\]%
\emph{and}
\[
A\,\Proj_w(\zeta)=Aw+K_w(\zeta-Hw).
\tag{\ref{eq:projaction}}
\]%

\begin{proof}
Problem \eqref{eq:proj} minimizes a strictly convex quadratic over an affine set, so a feasible point that satisfies the stationarity condition of the Lagrangian is its unique minimizer.

Let $\nu\in\R$ be a multiplier for the constraint $\one^\top g=1$ and $\theta\in\R^{n_z}$ a multiplier for the constraint $Hg=\zeta$. Then the Lagrangian is:
\[
\mathcal L(g,\nu,\theta)\coloneqq\frac12\sum_i\frac{g_i^2}{w_i}-\nu\,(\one^\top g-1)-\theta^\top(Hg-\zeta),
\]
where the factor $\frac12$ does not change the minimizer. Since $Hg=\sum_ig_iz_i$, the derivative of $\mathcal L$ with respect to $g_i$ is $g_i/w_i-\nu-\theta^\top z_i$. Setting it to zero for every $i$ gives $g_i=w_i\,(\nu+z_i^\top\theta)$, or in vector form:
\[
g=\mathrm{diag}(w)\,(\nu\one+H^\top\theta).
\]
It remains to choose $\nu$ and $\theta$ such that $g$ is feasible.

\emph{Rewriting the weights.} To separate the two constraints, we use $\mathrm{diag}(w)=W_w+ww^\top$, which is the definition of $W_w$ in \eqref{eq:moments}, together with $\mathrm{diag}(w)\one=w$:
\[
g=\nu w+W_wH^\top\theta+w\,(w^\top H^\top\theta)=\nu'w+W_wH^\top\theta,\qquad\nu'\coloneqq\nu+w^\top H^\top\theta.
\]
The benefit of this form is that $\one^\top W_w=w^\top-w^\top=0$, so the term $W_wH^\top\theta$ does not affect the sum of the weights.

\emph{Imposing the constraints.} For the first constraint, $\one^\top g=\nu'\,\one^\top w+\one^\top W_wH^\top\theta=\nu'$, so $\one^\top g=1$ fixes $\nu'=1$. For the second, $Hg=Hw+HW_wH^\top\theta=Hw+S_w\theta$, so $Hg=\zeta$ implies $S_w\theta=\zeta-Hw$. Since $S_w\succ0$ is invertible, $\theta=S_w^{-1}(\zeta-Hw)$. Inserting $\nu'=1$ and this $\theta$ gives
\[
g=w+W_wH^\top S_w^{-1}(\zeta-Hw),
\]
which is feasible and satisfies the stationarity condition, and is therefore the unique minimizer $\Proj_w(\zeta)$.

\emph{The action.} Multiplying by $A$ and using $AW_wH^\top S_w^{-1}=K_w$ from \eqref{eq:moments} gives \eqref{eq:projaction}.
\end{proof}

\section{Details for Section~\ref{sec:method}}\label{app:method}

\subsection{Detailed Implementation}\label{app:alg}

Algorithm~\ref{alg:full} records the full \mthd{BD-$\chi^2$} algorithm.
In the implementation, we add $\varepsilon\I$, with $\varepsilon=10^{-6}$, to covariance matrices before inversion for numerical stability.
At each query, we keep the $k$ nearest histories. 
The prior is uniform on the selected windows. 

\begin{algorithm2e}[h]
\caption{\mthd{BD-$\chi^2$}: one control step}
\label{alg:full}
\DontPrintSemicolon
\LinesNumbered
\SetKwInOut{Input}{Input}
\SetKwInOut{Parameters}{Parameters}
\SetKwInOut{Optional}{Optional}
\SetKwInOut{Output}{Output}
\SetKwFunction{FitLine}{FitLine}
\SetKwFunction{Sample}{Sample}
\SetKwProg{Fn}{function}{}{}
\SetKwComment{Comment}{$\triangleright$\ }{}
\SetCommentSty{algcmt}
\Input{windows $(z_i,a_i)_{i=1}^N$, prior $\pi$\\
latent $\xi\sim\N(0,\I)$, drawn once per episode\\
current history $z$}
\Parameters{$\tau,\sigma_m,r,T_{\mathrm{it}},\mu_{\min}, k$}
\Optional{line $(z^-,c^-,K^-)$ stored at the previous step}
\Output{action $u$}
{Keep the $k$ windows nearest to $z$ and use only these windows below.}\;
$q^\tau_i\gets\pi_i\,e^{-\|z_i-z\|^2/2\tau^2}\big/\textstyle\sum_j\pi_j\,e^{-\|z_j-z\|^2/2\tau^2}$
  \Comment*[r]{windows with similar histories}
\If(\Comment*[f]{{keep the current line}}){a line is stored}{
  $(c,K,w,\mu)\gets$ \FitLine{$c^-+K^-(z-z^-),\,K^-$}
    \Comment*[r]{carry line to $z$, refit}
  \lIf(\Comment*[f]{{line has left the data}}){$\mu<\mu_{\min}$}{discard the line}
}
\If(\Comment*[f]{{choose a starting point}}){no line is kept}{
  $a^\smp\gets$ \Sample{$\xi$}\;
  $(c,K,w,\mu)\gets$ \FitLine{$a^\smp,\,0$}
    \Comment*[r]{flat line through the sample}
}
$\zb\gets\sum_iw_iz_i$,\ \ $S\gets\sum_iw_i(z_i-\zb)(z_i-\zb)^\top$\;
$\alpha\gets\min\big\{1,\ r\big/\big(\|S^{-1/2}(z-\zb)\|\sqrt{\textstyle\sum_iw_i^2}\big)\big\}$
  \Comment*[r]{follow the line at most $r\sqrt{n_\eff}$ std.\ devs.}
$u\gets c-(1-\alpha)K(z-\zb)$
  \Comment*[r]{$\alpha=1$: the line at $z$; $\alpha=0$: mean action}
store $(z,c,K)$; \KwRet{$u$}\;
\BlankLine
\Fn(\Comment*[f]{refit the line}){\FitLine{$c,K$}}{
  \For{$j=1,\dots,T_{\mathrm{it}}$}{
    $e_i\gets a_i-c-K(z_i-z)$ \Comment*[r]{residual of window $i$ from the line}
    $w_i\gets q^\tau_i\,e^{-\|e_i\|^2/2\sigma_m^2}\big/\textstyle\sum_jq^\tau_j\,e^{-\|e_j\|^2/2\sigma_m^2}$
      \Comment*[r]{windows that follow the line}
    $(c,K)\gets\argmin_{c',K'}\sum_iw_i\|a_i-c'-K'(z_i-z)\|^2$
      \Comment*[r]{weighted least squares}
  }
  $\mu\gets\sum_iq^\tau_i\,e^{-\|a_i-c-K(z_i-z)\|^2/2\sigma_m^2}$
    \Comment*[r]{share of kernel weight on the line}
  \KwRet{$(c,K,w,\mu)$} \Comment*[r]{$w$: weights of the last refit}
}
\BlankLine
\Fn(\Comment*[f]{closed-form diffusion, $z$ fixed}){\Sample{$\xi$}}{
  $\tilde a\gets\sigma_0\xi$ \Comment*[r]{noise levels $\sigma_0>\sigma_1>\dots>\sigma_S=0$}
  \For{$k=0,\dots,S-1$}{
    $g^{\sigma_k}_i\gets q^\tau_i\,e^{-\|a_i-\tilde a\|^2/2\sigma_k^2}\big/\textstyle\sum_jq^\tau_j\,e^{-\|a_j-\tilde a\|^2/2\sigma_k^2}$
      \Comment*[r]{posterior over windows}
    $\tilde a\gets\tilde a+(\sigma_{k+1}-\sigma_k)\big(\tilde a-\sum_ig^{\sigma_k}_ia_i\big)/\sigma_k$
      \Comment*[r]{Euler step of \eqref{eq:ode}}
  }
  \KwRet{$\tilde a$} \Comment*[r]{{close to a stored action}}
}
\end{algorithm2e}

\clearpage
\section{Details for Control Benchmark Setup}\label{app:tasks}

This section describes the three tasks of Section~\ref{sec:control_benchmark}, which we took from \citet{naf2025choose}. In all tasks, the data are collected by an MPC expert with zero-mean Gaussian noise added to its inputs, and no noise is added during evaluation. We collect demonstrations using custom MPC experts: nonlinear MPC implemented with CasADi \citep{andersson2019casadi} and solved using IPOPT \citep{wachter2006implementation} for Cartpole and Reacher, and linear MPC implemented with CVXPY \citep{diamond2016cvxpy,agrawal2018rewriting} and solved using OSQP \citep{stellato2020osqp} for Rocket Landing.
The reported cost is a quadratic penalty on the quantities listed below plus an input penalty, accumulated over the episode and divided by the expert's cost from the same initial state.

\paragraph{Cart-pole Swing-up.} The observation is $y=(p,\sin\theta,\cos\theta,\dot p,\dot\theta)$, where $p$ is the cart position and $\theta$ is the pole angle, with $\theta=0$ upright and $\theta=\pi$ hanging down. The input $u$ is the actuator command, which applies a horizontal force to the cart, with $|u|\le3$. The data contain 200 trajectories of 100 steps at 25\,Hz, collected by an MPC expert with a horizon of 100 steps.
Each test episode runs for 300 steps. The cart position is drawn uniformly from $[-0.1,0.1]$\,m, the pole angle uniformly within $0.1$\,rad of the hanging position, and both velocities uniformly from $[0.001,0.02]$ (in m/s and rad/s). A trial is successful if, throughout the final 50 steps, $|\theta|\le15^\circ$, $|\dot\theta|\le1$\,rad/s and $|p|\le1$\,m. The cost penalizes the deviation from the upright equilibrium.

\paragraph{Rocket Landing.} The observation is $y=(p_x,p_y,v_x,v_y,\theta,\dot\theta)$, the position, velocity, tilt and angular velocity of a planar rocket. The input $u$ contains normalized commands for the main thrust ($0\le u_1\le1$), the side thrust, and the nozzle angle ($|u_2|,|u_3|\le1$). The expert uses the simulator's linearization about hover, with a horizon of 60 steps at 60\,Hz. 
The data contain the 58 of the MPC's 60 episodes that landed without crashing. The noise added to the expert's inputs has standard deviations $(0.05,0.1,0.1)$. Each test episode runs for at most 1000 steps (about 16.7\,s) and starts with the rocket at rest. One start is the fixed start of the original example of \citet{naf2025choose}. The other nine are drawn with seeds not used for data collection, at random positions in the upper part of the environment and with tilts of up to $17.5^\circ$. A trial is successful if the rocket lands on the pad and comes to rest. The episode ends early if the rocket leaves the environment, tilts beyond about $41^\circ$, or crashes. The cost penalizes the deviation from a reference point.

\paragraph{Reacher.} The observation is $y=(\cos q,\sin q,p_x,p_y,\dot q)$, where $q=(q_1,q_2)$ are the angles of the arm's two joints, $(p_x,p_y)$ is the end effector position, and $\cos q$ and $\sin q$ are taken elementwise. The input $u$ contains the commands to the two joint motors, with $|u_i|\le0.5$. The end effector must reach five reference positions in a given order, each to within $5$\,mm, and the next reference becomes active once one is reached. For our methods, the active reference is appended to the observation. The data contain 200 trajectories of 200 steps at 50\,Hz, each following a random sequence of five references, collected by an MPC expert with a horizon of 30 steps. The noise added to the expert's inputs has standard deviation $0.05$. Each test episode starts with the arm nearly straight and at rest, and ends when the fifth reference is reached or after 500 steps (10\,s). Five test episodes use the reference sequence of the original example of \citet{naf2025choose}, and five use random sequences, drawn, like all 10 initial states, with seeds not used for data collection. A trial is successful if the end effector reaches all five references in order while staying below $p_y=0.1$\,m. Violations of this height limit do not end the episode. The cost penalizes the distance to the active reference and the joint velocities.
Counting reference completion alone gives $10/10$ for \mthd{BD-$\chi^2$} and $5/10$ for \mthd{BD-mean-$\chi^2$}.

\section{Control Benchmark Results}\label{app:control_results}
\begin{figure}[h]
\centering
\includegraphics[width=0.9\textwidth]{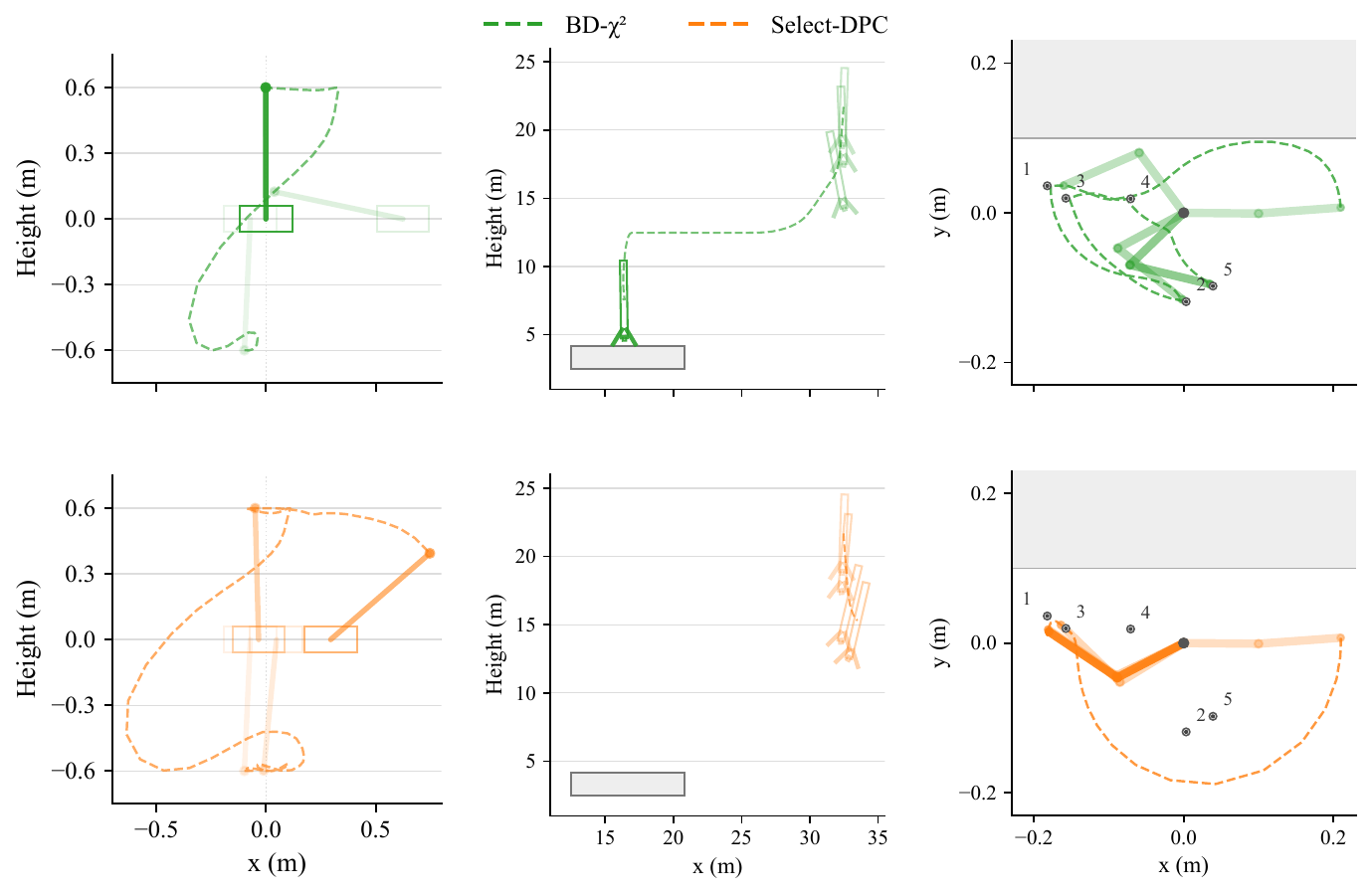}
\caption{Control benchmark examples where \mthd{Select-DPC} fails. Dashed lines denote the trajectories followed.}
\label{fig:trajectory_comparison}
\end{figure}

\begin{figure}[h]
\centering
\includegraphics[width=0.8\textwidth]{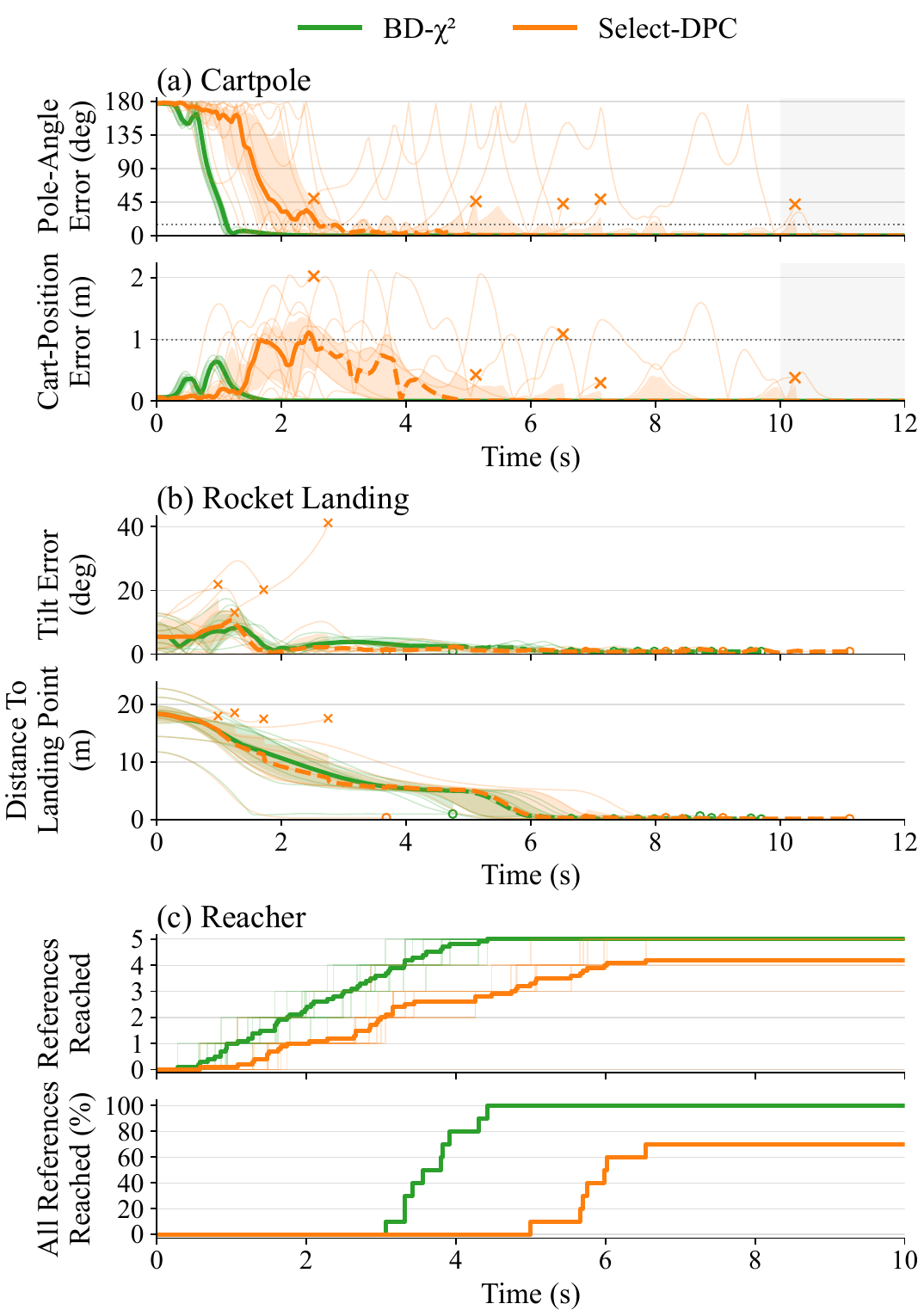}
\caption{Closed-loop behavior on the control benchmarks across the 10 initial states. (a) Cart-pole: tracking errors. (b) Rocket landing: tracking errors. (c) Reacher: reference completion. In (a) and (b), thin lines show individual trials, thick lines the median, and bands the interquartile range. $\circ$ marks a successful run and $\times$ an early termination due to a violation of the state constraints. After a trial ends, its values are omitted, and dashed medians and bands summarize only the remaining trials.}
\label{fig:control_tracking}
\end{figure}

Table~\ref{tab:results} reports the results. \mthd{BD-$\chi^2$} and \mthd{BD-mean-$\chi^2$} succeed most often, in 27 and 24 of the 30 trials. The policies of Section~\ref{sec:limits} rarely succeed. \mthd{BD-mean} and \mthd{BD-sample} make progress but do not stabilize: they swing the pole up but do not keep it upright, overshoot or undershoot the landing pad, and move jerkily on the reacher, where \mthd{BD-sample} reaches more references than \mthd{BD-mean}. \mthd{BD-KL} returns no action on any start, because its constraint set is empty at the initial history (Section~\ref{sec:constraint}).
On the cart-pole, \mthd{Select-DPC} can lose the pole shortly after swinging it up (Fig.~\ref{fig:trajectory_comparison} left). On rocket landing, all of its failures are violations of the state constraints (Fig.~\ref{fig:trajectory_comparison} middle). On the reacher, its three failures come from freezing before all references are reached (Fig.~\ref{fig:trajectory_comparison} right). 
The three failures of \mthd{BD-$\chi^2$} on the reacher reach all references in order but exceed the height limit, with a maximum violation of about $1$\,cm. Five failures of \mthd{BD-mean-$\chi^2$} on the reacher reach the first few references and then freeze, and the sixth reaches all five references but exceeds the height limit by $0.2$\,mm.
\mthd{DeePC} succeeds only once on the cart-pole and on rocket landing. It rarely swings the pole up and, when it does, does not keep it upright. \mthd{Windowed-DPC} rarely swings the pole up and often moves away from the landing target.
Plots of tracking errors for cart-pole and rocket landing can be found in Figure \ref{fig:control_tracking} (a) and Figure \ref{fig:control_tracking} (b), and plot of reference tracking progress of the reacher task can be found in Figure \ref{fig:control_tracking} (c).

In terms of computation time, \mthd{BD-$\chi^2$} takes between $0.3$ and $0.6$\,ms to compute each action. The baselines solve an optimization problem at every step and take from $73$\,ms (\mthd{Select-DPC} on the cart-pole) to $54$\,s (\mthd{DeePC} on rocket landing).

\clearpage
\section{Push-T Details}\label{app:pusht}

\indent
{\bf Simulator and data.} We use the Push-T environment of \citet{chi2023diffusion}. The pusher tracks the commanded position with a PD controller simulated at 100\,Hz, and the inputs are applied at 10\,Hz, so an episode of 300 steps lasts 30\,s. The workspace is $512\times512$ pixels, and the target pose is fixed at its center with a $45^\circ$ rotation. The 206 released demonstrations give $25{,}444$ observation--input pairs. An initial state places the pusher uniformly in $[50,450]^2$ and the block uniformly in $[100,400]^2$ (in pixels), with a normally distributed block angle.

\noindent
{\bf Methods.} Both methods use the last two observations as their history. They compute a sequence of inputs and apply the first ones before computing the next sequence. \mthd{BD-$\chi^2$} uses windows whose action is the next inputs of the demonstration (the action-chunk variant of Section~\ref{sec:setup}). \mthd{Select-DPC} plans over its horizon, with a cost that weights the standardized block pose $(x,y,\cos\theta,\sin\theta)$ by $100$ and the inputs by $0.01$.

\noindent
{\bf Tuning.} Tuning proceeds in two stages, both on the same 8 validation initial states. To keep the cost of \mthd{Select-DPC} manageable, the first stage evaluates every setting with a fixed random subset of 64 demonstrations.
For \mthd{BD-$\chi^2$}, the settings are 8 or 16 predicted inputs, 4 or 8 applied inputs, $\tau\in\{0.05,0.1,0.2\}$ and $\sigma_m\in\{0.1,0.3,1\}$ (36 settings). 
For \mthd{Select-DPC}, we test all combinations of horizons 8 or 16, 4 or 8 applied inputs, $l_1$ weights $\{10, 100, 1000\}$, and $\{200, 500, 1000\}$ selected windows, which gives 36 settings. We also test horizon 30 with 8 applied inputs and $l_1$ weight 100 at each of the three window counts, yielding 39 settings in total.
The second stage reevaluates the two best settings of each method, together with its initial setting, using all demonstrations. \mthd{BD-$\chi^2$}, whose diffusion sample is random, is evaluated with three draws per initial state. Settings are ranked by mean score, then by success rate.
The selected settings are 8 predicted and 8 applied inputs, $\tau=0.1$ and $\sigma_m=0.3$ for \mthd{BD-$\chi^2$}, and a horizon of 30, 8 applied inputs, $\ell_1$ weight $100$ and 200 selected windows for \mthd{Select-DPC}.

\noindent
{\bf Testing.} The 50 test initial states are drawn with seeds not used for tuning, and each is run once per method.

\begin{figure}[h]
\centering
\includegraphics[width=\textwidth]{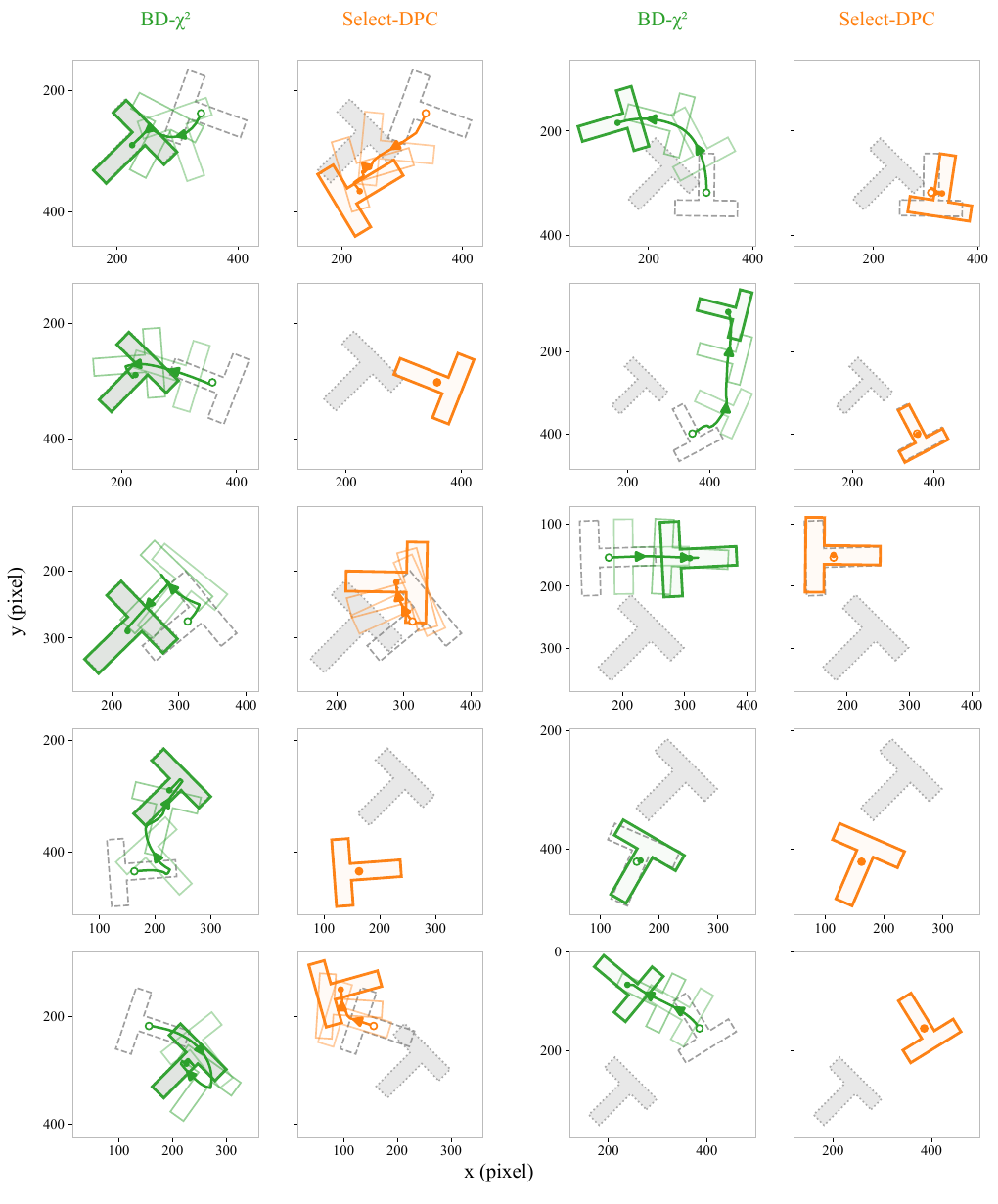}
\caption{Push-T trajectories under \mthd{BD-$\chi^2$} (green) and \mthd{Select-DPC} (orange). Solid curves trace the block's center of mass, with arrows indicating motion direction. Gray dashed outlines mark initial poses, and gray filled shapes with dotted borders show reference poses. Faint and bold colored outlines indicate intermediate and final poses, respectively. Open and filled circles mark trajectory starts and ends. Left pairs illustrate \mthd{BD-$\chi^2$} successes and right pairs illustrate failures.}
\label{fig:pusht_trajectory_comparison}
\end{figure}

\end{document}